\documentclass[pre,fleqn,showpacs]{revtex4}
\usepackage{graphicx, amsmath, wrapfig,amsfonts,amssymb,mciteplus,bm}
\usepackage[dvips]{color}
\newcommand{\meanv}[1]{{\left\langle #1 \right\rangle}}
\newcommand{\vv}[1]{{\bm #1}}
\newcommand{\mat}[1]{{\sf #1}}
\newcommand{\dd}{\partial}

\renewcommand{\a}{\alpha}
\renewcommand{\b}{\beta}
\renewcommand{\c}{\gamma}
\renewcommand{\d}{\delta}
\renewcommand{\t}{\theta}
\renewcommand{\r}{\rho}

\def\htt{\hat{\theta}}
\def\hr{\hat{\rho}}
\def\hg{\hat{g}}
\def\hp{\hat{\phi}}
\def\hn{\hat{\nu}}

\def\hvg{\hat{\vv{g}}}
\def\vg{\vv{g}}
\def\vr{\vv{r}}
\def\vn{\vv{\nu}}
\def\hvn{\hat{\vv{\nu}}}
\def\dr{\delta \rho}

\def\vk{\vv{k}}

\def\vq{\vv{q}}
\def\vp{\vv{p}}
\def\vphi{\vv{\phi}}
\def\vpsi{\vv{\psi}}

\begin{document}
\title{A FDR-preserving field theory of glass transition in terms of the
fluctuating hydrodynamics}
\author{Taka H. Nishino}
\email{takahiro@yukawa.kyoto-u.ac.jp}
\author{Hisao Hayakawa}
\email{hisao@yukawa.kyoto-u.ac.jp}
\affiliation{Yukawa Institute for Theoretical Physics, Kyoto University,\\
Kitashirakawa-Oiwakecho, Sakyo-ku, Kyoto, 606-8502, Japan.}
\date{\today}
\begin{abstract}
 A field theoretical method for the fluctuating hydrodynamics
 with preserving fluctuation-dissipation relations (FDR)
 is reformulated.
 It is shown that the long time behavior within the first-loop order
 perturbation under the assumption
 that the correlations include the momentum decay fast enough,
 is equivalent to that for the standard mode-coupling theory.
\end{abstract}

\pacs{64.70.Q-, 61.20.Lc, 05.40.-a, 03.50.-z}
\maketitle

\section{Introduction}\label{sINTRO}

In the vicinity of the glass transition point,
the dynamics of supercooled liquids
becomes extremely slow~\cite{review-ediger,review-angell,review-debenedetti}.
The dynamics of the glass transition attracted much attention over the years.
Among many theoretical approaches, the mode-coupling theory (MCT)
is one of the most successful ones
which can be ``derived'' from the first principle equation,
and explains many aspects of observations in  experiments and simulations, 
such as multi-step relaxation processes and
the Debye-Waller
parameter~\cite{review-gotze,review-das,review-reichman,review-miyazaki}.

In spite of such advantages of MCT, there are some
controversial points for the validity of the
standard MCT (SMCT).
Indeed, SMCT predicts the existence of
the ergodic-nonergodic (ENE) transition,
where the system becomes nonergodic
below a critical temperature or above a critical density,
 while real systems
are still ergodic in experiments and simulations at low temperature
or high density.
Furthermore, SMCT predicts an algebraic divergence of the viscosity
at the critical point of ENE transition,
but the viscosity for real supercooled liquids obeys the
Vogel-Fulcher law near the glass transition point and the
Vogel-Fulcher temperature is lower than the
critical temperature of ENE transition.
To overcome these difficulties of SMCT,
many investigations have been carried out~\cite{
dm,gotze1987,sdd,kawasaki1994,kawasaki1997,
yamamoto1998,fuchizaki1998,kawasaki1998,*kawasaki2000,marconi1999,*marconi2000,
franz2000,szamel2003,
wu2005,miyazaki2005,
mayer2006,cates2006,abl,mazenko2006,*mazenko2007,biroli2006,
biroli2007,beither2007,*beither2007-2,kk,*kk2,szamel2007}.
The failures of SMCT may be originated from the decoupling approximation
of a four-point correlation function.
In fact, Mayer {\it et al.}~\cite{mayer2006}, introduced a toy model 
which does not have any spatial degree of freedom,
and demonstrated that the ergodicity of the system at the low temperature
is recovered when they include higher-order correlations, while 
there exists ENE transition within the framework of the decoupling approximation.
It suggests that we should not adopt the decoupling approximation, 
but use an approximation which contains higher-order correlations.
However, the systematic improvement of the approximation is difficult
within the conventional framework
by using the projection operator technique.

The field theoretical approach is
the promising method which can
systematically improve the approximations. 
There is another advantage of the field theory in which we can discuss
the response function and
the fluctuation-dissipation relations (FDR).
Following Martin-Siggia-Rose (MSR) method~\cite{msr}, 
we can construct an action by the introduction of conjugate fields, 
for a set of nonlinear Langevin equations, and can use the perturbative
expansion.
The current situation for this approach, however, is confusing. 
Indeed, among many field theoretical
investigations~\cite{dm,sdd,kawasaki1997,miyazaki2005,abl,kk,beither2007,szamel2007},
only a few papers have succeeded to derive SMCT in the lowest order
perturbation from the nonlinear Langevin equations.
One of main difficulties lies in the violation of FDR 
in each order of naive perturbative expansions of
the set of nonlinear Langevin equations,
as indicated by Miyazaki and Reichman~\cite{miyazaki2005}. 

In order to recover FDR-preservation at each order of
the perturbation, recently,
Andreanov, Biroli and Lef{\` e}vre (ABL)~\cite{abl}
indicated the importance of 
time reversal symmetry of the action, and introduced 
some additional field variables.
Indeed, ABL demonstrated that we can construct a
FDR-preserving field theory, starting 
from the nonlinear Langevin equations which contain both
the Dean-Kawasaki equation and the fluctuating nonlinear hydrodynamic (FNH) equations.
Kim and Kawasaki~\cite{kk} further improved ABL method
and they derived a mode coupling equation, similar to
SMCT, from Dean-Kawasaki equation~\cite{dean1996,kawasaki1994}
in the first-loop order via the irreducible memory functional approach
which to be essential for treating the dynamics of the
dissipative systems such as the interacting Brownian particle system.

On the other hand,
Das and Mazenko~\cite{dm} published a pioneer
paper on the field theoretic approach of FNH.
They suggested the existence of the cutoff mechanism in which the
system is always ergodic even at low temperature.
Later, Schmitz, Dufty and De (SDD)~\cite{sdd}
reached the same conclusion as that by Das and Mazenko
from  a concise discussion,
though they destroyed the Galilean invariance of FNH equations.
On the other hand,
Kawasaki~\cite{kawasaki1994} suggested that FNH equations
reduce to Dean-Kawasaki equation in the long time limit.
Furthermore, ABL~\cite{abl} suggested the existence of
ENE transition in FNH,
and indicated that the calculation by Das and Mazenko breaks FDR-preservation.
Moreover, Cates and Ramaswamy~\cite{cates2006} indicated
that the calculation by Das and Mazenko violates the momentum
preservation.
Das and Mazenko~\cite{dm2}, however, responded that
the indications by ABL and by Cates and Ramaswamy
are not the fatal errors of Das and Mazenko~\cite{dm}, but contain some misleading arguments.
Thus, we are still in  a confusing situation for the application
of the field theory to the glass transition, 
and cannot conclude whether ENE transition exists in FNH.

In this paper,
we apply the method developed by Kim and Kawasaki~\cite{kk} to FNH 
to clarify the current situation
of the FDR-preserving field theoretical approach to the glass transition.
The organization of this paper is as follows.
In the next section, we introduce FNH which
describes the time evolutions of the density field and the momentum
field,
agitated by the fluctuating random force,
for  compressible fluids.
This set of equations is equivalent to that
used by Das and Mazenko~\cite{dm} and ABL~\cite{abl}.
In the former half of Section III,
we make an action invariant under the time-reversal transformation.
In order to keep the linearity of the time-reversal transformation, we introduce
some additional variables and their conjugate fields.
This linearity of the time-reversal transformation makes
FDR-preserving field theory possible.
We also introduce a complete set of Schwinger-Dyson equations
of our problem, and summarize some identities used for the perturbative calculation in the latter half of Section III. 
Section~\ref{sFIRSTLOOP}
is the main part of our paper, in which we explain the detailed calculations
of perturbative expansion within the first-loop order,
under the assumption that the correlations including momentum
can be ignored in the long time limit.
Within this approximation, we predict the existence of ENE transition,
and reach an equivalent equation obtained from SMCT.
In the last section, we discuss the validity of our assumptions
used in this paper, and compare our results with
others.
We also summarize our results.
In Appendix A, we introduce the details of the time-reversal
transformation and some relevant relations derived from
the time-reversal transformation.
In Appendix B, we present the details of the
calculation for one component of the Schwinger-Dyson equation.
In Appendix C, we show some relations for the equal-time correlations and the self-energies.
In Appendix D, we write down the explicit expressions for
all three-point vertex functions.

\section{Fluctuating Nonlinear Hydrodynamics}\label{sFNH}

In this section, we briefly summarize our basic equations, FNH, 
and MSR action~\cite{msr}.
The argument in this section is parallel to those
in the previous studies~\cite{dm,abl}.

Let us describe
a system of supercooled liquids in terms of a set of equations for   
the density field $\r(\vv{r},t)$ and the momentum field $\vg(\vv{r},t)$.
For the continuity equation of momentum,
we employ Navier-Stokes equation for compressible fluids
supplemented with the osmotic pressure induced by the density
fluctuation,
and the noise caused by the fast fluctuations.
In order to keep the analysis simple, we ignore the
fluctuations of energy~\cite{kim1991} as assumed by
Das and Mazenko~\cite{dm}, SDD~\cite{sdd} and ABL~\cite{abl}.

The time evolutions of the collective variables $\rho$ and $\vg$,
which we call FNH equations, 
are given by~\cite{dm,abl} 
\begin{align}
 & \label{continue}
 \dd_t \r = - \nabla \cdot \vv{g} ,  \\
 &  \dd_t g_\a = - \rho \nabla_\a \frac{\delta F_U}{\delta \rho}
 - \nabla_\b \frac{g_\a g_\b}{\rho} - L_{\a\b} \frac{g_\b}{\rho} +
 \eta_\a. \label{navier}
\end{align}
Here, $\eta_\a$ is the Gaussian white noise with zero mean, 
which satisfies
\begin{equation}
 \label{mean}
 \meanv{\eta_\a\left(\vv{r},t\right) \eta_\b\left(\vv{r}',t'\right)}
  = 2T L_{\a\b} \delta \left(\vv{r}-\vv{r}'\right) \delta\left(t-t'\right),
\end{equation}
where $T$ is temperature
and $L_{\a\b}$ is the operator tensor
acting on any field variables $h \left(\vr\right)$ as
\begin{equation}
 L_{\a\b} h \left(\vv{r}\right)
  = - \left\{ \mu_0\left(\frac{1}{3}\nabla_\a\nabla_\b + \delta_{\a\b} \nabla^2\right)
  + \zeta_0 \nabla_\a\nabla_\b \right\} h\left(\vv{r}\right) ,
\end{equation}
with the shear viscosity $\mu_0$ and the bulk viscosity $\zeta_0$.
We note that here and after the Boltzmann's constant set to unity.
In this paper, the Greek indices, such as $\a$, are used for the spatial
components, and Einstein's rule such as
$g_\a g_\a \equiv \sum^3_{\a=1} g_\a^2$ is also adopted.
The effective free-energy functional $F=F_K+F_U$
consists of the kinetic part $F_K$ and the potential part $F_U$ as
\begin{align}
\label{f_kinetic}
 F_K = & \frac{1}{2} \int
 d\vv{r} \, \frac{\vv{g}^2 \left(\vr\right)}{\rho \left(\vr\right)} ,\\
 F_U = & \frac{T}{m} \int d\vv{r} \,
 \rho\left(\vv{r}\right) \left(\ln
 \left(\frac{\rho\left(\vv{r}\right)}{\rho_0} \right)
 -1 \right) 
 - \frac{T}{2 m^2} \int d\vv{r} d\vv{r}' \, c\left(\vv{r}-\vv{r}'\right)
 \delta\rho\left(\vv{r}\right) \delta\rho \left(\vv{r}'\right),
 \label{fu}
\end{align}
where 
$m$ is the mass of a particle 
and $c\left(\vr\right)$ is the direct correlation function~\cite{hansen}.
The potential part $F_U$ of the effective free-energy functional has the
same form with Ramakrishnan-Yussouff form~\cite{ramakrishnan_yussouff}.
Here,  $\dr\left(\vr,t\right) \equiv \r\left(\vr,t\right) - \r_0$
is the local density
fluctuation around the mean density $\r_0$.
From the relations (\ref{f_kinetic}) and (\ref{fu}),
we can rewrite \eqref{continue} and \eqref{navier} as
\begin{align}
 & \label{continue2}
  \dd_t \r = 
 -\nabla \left(\r \frac{\delta F}{\delta \vg} \right) , \\
 & 
 \dd_t g_\a = -\r \nabla_\a \frac{\d F}{\d \r} -
 \nabla_\b \left(g_\a \frac{\d F}{\d g_\b} \right)
 - g_\b \nabla_\a \frac{\d F}{\d g_\b} - L_{\a\b} \frac{\d F}{\d g_\b}
 + \eta_\a, \label{navier2}
\end{align}
where we have used 
$\r \nabla_\a (\d F_K/\d \r) = - g_\b \nabla_\a (\d F/\d g_\b)$.

In general, it is impossible to solve the set of nonlinear
partial differential equations~\eqref{continue}-\eqref{fu}.
In this paper, we adopt MSR field theory~\cite{msr}.
Let us derive the MSR action.
Because the collective variables $\r$ and $\vg$
satisfy the dynamic equations \eqref{continue2} and \eqref{navier2},
the average of an observable $A[\r,\vg]$ is expressed as
\begin{eqnarray}\label{eq:9}
 \meanv{A} &=&
  \left\langle
   \int D\r' D\vg' \, A[\r',\vg']
  \delta(\r'-\r) \delta(\vg' - \vg) \right\rangle \nonumber \\
 &=& \int D\r D\vg \, J(\r,\vg) A[\r,\vg]
  \Biggl\langle
  \delta\left(\dd_t \r + \nabla \left(\r \frac{\d F}{\d
                                 \vg}\right)\right) \nonumber
  \\
 &&
  \times \prod_\a
  \delta\left(\dd_t g_\a + \r \nabla_\a \frac{\d F}{\d \r} +
  \nabla_\b \left(g_\a \frac{\d F}{\d g_\b} \right)
 + g_\b \nabla_\a \frac{\d F}{\d g_\b} + L_{\a\b} \frac{\d F}{\d g_\b}
 - \eta_\a\right) \Biggr\rangle , 
\end{eqnarray}
where $J(\r,\vg)$ is the Jacobian.
As written in Ref.~\cite{jensen1981}, 
the Jacobian $J(\r,\vg)$ can be independent of both $\r$ and $\vg$
when we employ the It{\^ o} interpretation.
When we replace the delta functions by 
the functional integrals of the conjugate fields $\hr$ and $\hg_\a$,
the average of $A$ in eq.~\eqref{eq:9} can be rewritten as
\begin{eqnarray}
 \meanv{A} &=&
\frac{1}{Z_0}
 \int D\r D\vg D\hr D\hvg \,
  A[\r,\vg] \nonumber \\
 &&
 \times
 \Biggl\langle
  \exp\Biggl[
  \int d\vr dt \, \biggl\{- \hr
\left(
\dd_t \r + \nabla \left( \r \frac{\d F}{\d \vg}
                   \right)
\right) \nonumber\\
& &
  - \hg_\a 
\left(\dd_t g_\a + \r \nabla_\a \frac{\d F}{\d \r} +
  \nabla_\b \left(g_\a \frac{\d F}{\d g_\b} \right)
 + g_\b \nabla_\a \frac{\d F}{\d g_\b} + L_{\a\b} \frac{\d F}{\d g_\b}
 - \eta_\a
\right)
\biggr\}
\Biggr]  \Biggr\rangle ,
\end{eqnarray}
where $Z_0$ is the normalization constant.
By means of eq.~\eqref{mean}, the average of $A$ is given by 
\begin{eqnarray}
 \meanv{A} &=& 
\frac{1}{Z_0}
 \int D\r D\vg D\hr D\hvg \,
  A[\r,\vg] \nonumber \\
 &&
  \times
  \exp\left[\int d\vr dt \, 
\left\{
- \hr
\left(\dd_t \r + \nabla (\r \frac{\d F}{\d \vg})
\right)
  - \hg_\a \left(
\dd_t g_\a + \r \nabla_\a \frac{\d F}{\d \r} +
  \nabla_\b \left(g_\a \frac{\d F}{\d g_\b} \right)
 + g_\b \nabla_\a \frac{\d F}{\d g_\b} + L_{\a\b} \frac{\d F}{\d g_\b}
 \right)\right\}\right] \nonumber \\
 && \times \left\langle \exp\left[\int d\vr dt \, \hg_\a \eta_\a\right] \right\rangle  \nonumber \\
 &=& 
\frac{1}{Z_0}
 \int D\r D\vg D\hr D\hvg \,
  A[\r,\vg] e^{S[\rho, \hr,\vg,\hvg]} ,
\end{eqnarray}
where the MSR action $S[\rho, \hr,\vg,\hvg]$ is defined by
\begin{eqnarray}
  S[\rho, \hr,\vg,\hvg]
  &\equiv & \int d\vr dt \,
  \Biggl[- \hr \left\{
  \dd_t \rho +
  \nabla_\a \left(\rho \frac{\d F}{\d g_\a}\right) \right\}
  + T \hat{g}_\a L_{\a\b} \hat{g}_\b
 \nonumber \\
 && 
- \hat{g}_\a \biggl\{
  \dd_t g_\a
  + \rho \nabla_\a \frac{\delta F}{\delta \rho}+
  \nabla_\b  \left(g_\a \frac{\d F}{\d g_\b}\right)
  + g_\b \nabla_\a \frac{\d F}{\d g_\b}
  + L_{\a\b}\frac{\d F}{\d g_\b}
  \biggr\}
  \Biggr] .
  \label{action1}
\end{eqnarray}
%

\section{The construction of a FDR-preserving field theory}\label{sFDR}

\subsection{The time-reversal symmetry in the action}

In order to construct a FDR-preserving field theory,
it is necessary to introduce some new variables, and the linear time reversal transformation, 
which makes the MSR action invariant.
It is easy to check that the action~\eqref{action1}
is invariant under the time reversal transformation~\cite{abl}
\begin{equation}
 \label{reversal1}
 \begin{split}
  & t \to -t, \quad
  \r \to \r , \quad \hr \to -\hr + \frac{1}{T}\frac{\d F}{\d \r}, \\
  & g_\a \to -g_\a , \quad
  \hg_\a \to \hg_\a - \frac{1}{T} \frac{\d F}{\d g_\a}.
 \end{split}
\end{equation}
Here, we adopt the method developed by
Kim and Kawasaki~\cite{kk} in which
they introduced the new variable 
\begin{equation}
 \label{kkt}
 \theta_{KK} \equiv \frac{\d F_U}{\d \r} - A* \dr,
\end{equation}
where $A*\dr$ represents the linear part of ${\d F_U}/{\d \r}$ on $\dr$.
Kim and Kawasaki~\cite{kk}
confirmed that the density correlation function
of Dean-Kawasaki equation for the non-interacting case satisfies the
diffusion equation {\it nonperturbatively}.
Furthermore, they concluded that the nonergodic parameter is same
as that of SMCT.

Following the idea of Kim and Kawasaki~\cite{kk},
to eliminate the nonlinearity of time reversal transformation of
\eqref{reversal1}, we introduce the new variables $\theta$  and $\vn$ 
\begin{align}
 \label{myt}
 \theta & \equiv \frac{1}{T} \frac{\d F}{\d \r} - K*\dr,
 \\
 \label{mynu}
 \nu_\a & \equiv \frac{1}{T} \frac{\d F}{\d g_\a} - \frac{1}{T\r_0} g_\a, 
\end{align}
where the operator $K$ acts on any function $h$ as
\begin{equation}\label{coupling}
 K * h\left(\vr\right) \equiv \frac{1}{m\r_0} \int d\vr' \,
  \left\{  \delta\left(\vr-\vr'\right) - \frac{\r_0}{m} c\left(\vr - \vr'\right)
\right\}
  h\left(\vr'\right).
\end{equation}
It should be noted that the right-hand side (RHS) of
eqs.~\eqref{myt} and \eqref{mynu} do not include
the zeroth and the first order of $\dr$ and $\vg$.
The choices of eqs.~\eqref{myt} and \eqref{mynu}
differ from those by ABL~\cite{abl}.
The implication of the difference will be discussed 
in Section~\ref{sDIS}.

As the result of the introduction of $\theta$ and $\nu_\a$,
the action~\eqref{action1} can be rewritten as
\begin{align}
 S[\vpsi] 
 = &
 \int d\vr dt \, \Biggl[
  -\hr \left\{ \dd_t \r + \nabla_\a
  \left(\r \left(\r_0^{-1} g_\a + T \nu_\a \right)\right) \right\}
 - \hg_\a \biggl\{
 \dd_t g_\a + T \r \nabla_\a \left(K * \d \r + \theta\right)
 + L_{\a\b}
 \left(\r_0^{-1} g_\b + T \nu_\b\right)
 \nonumber \\ &
 + \nabla_\b(g_\a (\r_0^{-1} g_\b + T\nu_\b))
 + g_\b \nabla_\a (\r_0^{-1} g_\b + T\nu_\b)
 \biggr\}
 + T \hg_\a L_{\a\b} \hg_\b
 - \htt (\t - f_\t )
 - \hn_\a \left(\nu_\a - f_{\nu_\a} \right)
 \Biggr], \label{action2}
\end{align}
where we have introduced
\begin{eqnarray}
 f_\t (\dr,\vg) 
& \equiv &
\frac{1}{T} \frac{\d F}{\d \r} - K*\d\r, \\
 f_{\nu_\a} (\dr,\vg) 
& \equiv & 
\frac{1}{T} \frac{\d F}{\d g_\a} -
  \frac{1}{T\r_0} g_\a.
\label{eq20}
\end{eqnarray}
We have also used 
the abbreviation of a set of the field variables
$\vpsi^T \equiv \left(\d\r,\hr,\t,\htt,\vg,\hvg,\vn,\hvn\right)$.
Here, the time reversal transformation
which makes the action~\eqref{action2}
invariant, is given by
\begin{eqnarray}
 &&
   t \to -t , \quad
  \r \to \r , \quad \hr \to -\hr + \t + K*\dr,
  \nonumber \\ &&
 g_\a \to -g_\a , \quad \hg_\a \to \hg_\a - \nu_\a
  - \frac{1}{T\r_0} g_\a , \quad \t \to \t ,
 \nonumber \\ &&
 \label{reversal2}
 \htt \to \htt + \dd_t \r ,\quad
  \nu_\a \to - \nu_\a , \quad
  \hn_\a \to - \hn_\a - \dd_t g_\a.
\end{eqnarray}
We, thus, can construct a FDR-preserving field theory,
due to the linearity of the time reversal transformation~\eqref{reversal2}.
As in the usual cases, let us split the action~\eqref{action2}
into the Gaussian part $S_{g}$
which represents bilinear terms of the field variables 
and the non-Gaussian part $S_{ng}$ as
\begin{equation}
 \label{ss}
 S[\vpsi] = S_{g} [\vpsi] + S_{ng} [\vpsi],
\end{equation}
where
\begin{eqnarray}
 S_{g} [\vpsi]
 &=&
 \int d\vr dt \, \Biggl\{
 - \hat\rho \left\{
 \dd_t \rho + \nabla_\a g_\a + \underline{T\r_0 \nabla_\a \nu_a } \right\}
 - \hat{g}_\a \biggl\{\dd_t g_\a + T\rho_0
 \nabla_\a K * \delta \rho +  T\rho_0 \nabla_\a \theta
 \nonumber \\ &&
 + L_{\a\b}
 \left(\r_0^{-1} g_\b + T \nu_\b\right) \biggr\}
 + T \hat{g}_\a L_{\a\b} \hat{g}_\b
 - \hat\theta \theta  - \hat\nu_\a\nu_\a
\Biggr\},
 \label{sg}
\end{eqnarray}
and
\begin{align}
S_{ng} [\vpsi] = &
 \int d\vr dt \,
 \Biggl\{-
 \hr  
 \left\{
 \underline{
 \nabla_\a \left(\d\r\left(\r_0^{-1} g_\a + T \nu_\a\right)\right)
 }
 \right\}
 - \hg_\a \Bigl\{ T \d\r \nabla_\a \left(K * \d\r+\t\right)
 \nonumber \\ &
 + \nabla_\b(g_\a (\r_0^{-1} g_\b + T\nu_\b))
 + g_\b \nabla_\a (\r_0^{-1} g_\b + T\nu_\b) 
 \Bigr\}
 + \htt f_\t(\dr,\vg)
 + 
\hn_\a f_{\nu_\a}(\dr,\vg)
\Biggr\}.
 \label{sng}
\end{align}
Note that we present some relations in  time-reversal symmetry of this action in Appendix A.

It should be noted that the continuity equation \eqref{continue2} 
can be rewritten as
\begin{eqnarray}
 \dd_t \r 
  &=& -\nabla_\a (\r (T\nu_\a + \r_0^{-1} g_\a)) \nonumber\\
 &=& - \nabla \cdot \vg - T\r_0 
\nabla_\a \cdot \nu_\a 
  - \nabla_\a (\d\r(T\nu_\a + \r_0^{-1} g_\a)),
  \label{continue9}
\end{eqnarray}
where we have used eq.~\eqref{mynu}.
From eqs.~\eqref{continue} and \eqref{continue9} we obtain the identity
\begin{equation}
 T\r_0 \nabla_\a \nu_\a + \nabla_\a (\d\r(T\nu_\a + \r_0^{-1} g_\a)) =
  0.
  \label{identity1}
\end{equation}
Therefore, the sum of 
the underlined terms in eqs.~\eqref{sg} and \eqref{sng} should be zero.
However, each the underlined term is included in the Gaussian
part~\eqref{sg} or the non-Gaussian part~\eqref{sng} 
 separately.
To satisfy the action invariant under the
time-reversal transformation in
`each' part, we should keep these terms in the calculation. 

\subsection{The exact results of  the Schwinger-Dyson equation}

In this subsection, we derive a set of closed equations
of two-point correlation function.
Let us express the two-point correlation
function in the matrix form as
\begin{equation}
 \mat{G} \left(\vr-\vr',t-t'\right) \equiv
  \meanv{\vpsi \left(\vr,t\right) \vpsi^T \left(\vr',t'\right)},
\end{equation}
and its $\psi \psi'$ component is represented by
\begin{equation}
 {G}_{\psi\psi'} \left(\vr-\vr',t-t'\right) \equiv
  \meanv{\psi \left(\vr,t\right) \psi' \left(\vr',t'\right)},
\end{equation}
where $\psi$ or $\psi'$ is the one of the components of $\vpsi$.
We note that here and after we adopt the simple notation $\psi=\rho$ to represent the contribution from $\delta\rho$.
With the aid of  the Fourier transform of
$h\left(\vr\right)$ 
\begin{equation}
 h\left(\vr\right) = \int \frac{d\vk}{\left(2\pi\right)^3} \,
  e^{i\vk\cdot\vr} h\left(\vk\right) ,
\end{equation}
and the action~\eqref{ss},
we obtain the Schwinger-Dyson (SD) equation  
\begin{equation}
 \label{sd}
  [\mat{G_0}^{-1} \cdot \mat{G} ]\left(\vk,t\right) -
  [\mat{\Sigma} \cdot \mat{G}] \left(\vk,t\right)
 = \mat{I} \, \delta\left(t\right),
\end{equation}
where $\mat{\Sigma}$ and $\mat{I}$ are, respectively,
the self-energy matrix and  the unit matrix.
Here, the free propagator matrix  $\mat{G_0}$ satisfies
\begin{equation}\label{S_g}
 S_g[\vpsi] = - \frac{1}{2}
  \int dX_1 dX_2 \, \vpsi^T (X_1) \mat{G_0}^{-1} (X_1 - X_2) \vpsi(X_2),
\end{equation}
where we have used the abbreviations as
$X_i \equiv \left(\vr_i,t_i\right)$ with $i=1,2$.
We note that the SD equation~\eqref{sd} is an equation for $16 \times 16$ matrices.

Here, we indicate that $G_{\r\r}(\vv{k},0)$ is related to 
the static structure factor $S\left(\vk\right)$ as
\begin{equation}
 \label{ssf}
 S\left(\vk\right) \equiv \frac{1}{m \r_0} G_{\r\r}\left(\vk,0\right).
\end{equation}
From the definition of the direct correlation function~\cite{hansen},
eq.~\eqref{ssf} can be rewritten as 
\begin{equation}\label{ssf2}
 S(\vk)= \left(1 - \frac{\r_0}{m} c\left(\vk\right)\right)^{-1} =
 \frac{1}{m\r_0} K^{-1}(\vk).
\end{equation}

Let us explicitly write some components of the SD equation 
\begin{equation}
 \label{kiso1}
  \dd_t G_{\r\phi} \left(\vk,t\right) + ik_\a
  G_{g_\a \phi}\left(\vk,t\right) + iT\r_0 k_\a G_{\nu_\a \phi}\left(\vk,t\right)
  = F_{\hr \phi} \left(\vk,t\right),
\end{equation}
\begin{equation}
 \label{kiso2}
G_{\t\phi} \left(\vk,t\right) + \Sigma_{\htt\htt}\left(\vk,0\right) G_{\r\phi}\left(\vk,t\right)
= F_{\htt \phi} \left(\vk,t\right),
\end{equation}
\begin{eqnarray}
\dd_t G_{g_\a \phi}\left(\vk,t\right)
  + T L_{\a\b} \left(\frac{1}{T\rho_0} G_{g_\b\phi}\left(\vk,t\right)
  + G_{\nu_\b\phi}\left(\vk,t\right) \right)
  + iT\r_0 k_\a \left(K\left(\vk\right) G_{\r\phi}\left(\vk,t\right) + G_{\t\phi}\left(\vk,t\right)\right)
  = F_{\hg_\a \phi} \left(\vk,t\right),
  \label{kiso3}
\end{eqnarray}
\begin{equation}
 \label{kiso4}
 G_{\nu_\a \phi} \left(\vk,t\right)
  +
  \Sigma_{\hn_\a\hn_\b}\left(\vk,0\right) G_{g_\b \phi}\left(\vk,t\right)
 = F_{\hn_\a \phi} \left(\vk,t\right),
\end{equation}
where $\phi$ is the one of the components of
$ \vphi^T \equiv \left(\d\r,\t,\vg,\vn\right)$ .
Here, eqs.~\eqref{kiso1}, \eqref{kiso2}, \eqref{kiso3}
and \eqref{kiso4} are obtained from $\hr\phi$, $\htt\phi$, $\hg_\a\phi$
and  $\hn_\a\phi$ components of the SD equation, respectively.  
The derivation of eq.~\eqref{kiso2} is shown in APPENDIX~\ref{csd} as
one example. The derivation of the other equations is parallel to that for eq.~\eqref{kiso2}.
We express the RHS of eqs.~\eqref{kiso1}-\eqref{kiso4} 
 as a unified form $F_{\hat{\phi} \phi}$ where 
$\hat\phi$ is the one of the components of
$\hat\vphi^T \equiv \left(\hr,\htt,\hvg,\hvn\right)$. 
From the parallel argument to derive $F_{\htt \phi}$, i.e. RHS of
(\ref{httr}), $F_{\hp\phi}$ satisfies
\begin{eqnarray}\label{eq37}
 F_{\hp \phi}(\vk,t) & =&
  \int^t_0 ds \,
 \biggl\{- \Sigma_{\hp \hr} \left(\vk,t-s\right) \left(K\left(\vk\right) G_{\r\phi}\left(\vk,s\right) + G_{\t\phi}\left(\vk,s\right)\right)
 + \Sigma_{\hp\htt}\left(\vk,t-s\right) \dd_s G_{\r\phi} \left(\vk,s\right)
 \nonumber \\  
& & - \Sigma_{\hp\hg_\a}\left(\vk,t-s\right)
 \left(\frac{1}{T\r_0}G_{g_\a\phi}\left(\vk,s\right)
   +
 G_{\nu_\a\phi}\left(\vk,s\right)\right)
   + \Sigma_{\hp\hn_\a} \left(\vk,t-s\right) \dd_s G_{g_\a
 \phi}\left(\vk,s\right) \biggr\}.
 \label{fff}
\end{eqnarray}

These components of SD equation are so complicated because of the self-energies.
However, we can simplify the components of the SD equation
with the aid of some exact relations.
First, we note that the equal-time correlation functions
satisfy (see~\eqref{C4_old} and \eqref{C9_old} in Appendix~\ref{scf})
\begin{equation}\label{zero1}
 \Sigma_{\htt\htt}(\vk,0) = 0, \quad {\rm and} \quad
\Sigma_{\hn_\a \hn_\b} (\vk,0) = 0. 
\end{equation}
With the aid of eqs.~\eqref{zero1} ,
eqs.~\eqref{kiso2} and \eqref{kiso4} are, respectively, simplified as
\begin{equation}
 \label{kiso2'}
  G_{\t\phi} \left(\vk,t\right) = F_{\htt \phi} \left(\vk,t\right),
\end{equation}
\begin{equation}
 \label{kiso4'}
 G_{\nu_\a \phi} \left(\vk,t\right) = F_{\hn_\a \phi} \left(\vk,t\right).
\end{equation}
Second, from  eqs.~\eqref{sg} and \eqref{sng}, there is 
the following identity
\begin{equation}
 \meanv{\frac{\d S[\psi]}{\d \hr(\vr,t)} \phi(\vr',t')} = 0.
  \label{exactr1}
\end{equation}
This identity can be expressed by the explicit form
\begin{eqnarray}
 \meanv{\left[\dd_t \r + \nabla \cdot \vg
        + T\r_0 \nabla \cdot \vn \cdot (\dr (T \vn + \r_0^{-1}\vg))
        \right] (\vr,t) \phi(\vr',t')} = 0.
\end{eqnarray}
With the aide of the identity~\eqref{identity1}, the Fourier transform
of this equation becomes
\begin{equation}
 \label{kiso1'}
 \dd_t G_{\r\phi}(\vk,t) + ik_\a G_{g_\a \phi} (\vk,t) = 0.
\end{equation}
This equation implies that our SD equation preserves the mass conservation law.
We also derive the identity by the substitution
of \eqref{kiso1'} from \eqref{kiso1} 
\begin{equation}\label{eq44}
 F_{\hr\psi} \left(\vk,t\right) = ik_\a T\r_0 F_{\hn_\a\psi} \left(\vk,t\right).
\end{equation}

\section{Perturbation in the first-loop order}\label{sFIRSTLOOP}

In this section,
we develop the perturbative calculation of the SD equation
within the first-loop order approximation.
When we assume that the correlations include the momentum decay fast enough,
we can obtain an equation for the non-ergodic parameter in the long time limit. 

From eqs.~\eqref{kiso3}, \eqref{kiso2'}, \eqref{kiso4'} and
\eqref{kiso1'},  we, thus, obtain the time
evolution of the density correlation function as
\begin{eqnarray}
 && \dd_t^2 G_{\r\r} \left(\vk,t\right)
 + \r_0^{-1} L \dd_t G_{\r\r} \left(\vk,t\right) + T\r_0 k^2
  K (\vk ) G_{\r\r}\left(\vk,t\right)
 \nonumber \\ 
 &=& - T\r_0 k^2 F_{\htt\r}\left(\vk,t\right)
   - ik_\a F_{\hg_\a \r}\left(\vk,t\right)
   + i T L k_\a F_{\hn_\a \r}\left(\vk,t\right),
   \label{ne}
\end{eqnarray} 
where $L \equiv \delta_{\a\b} L_{\a\b}$.
This is a remarkable result that the left-hand side
(LHS) of eq.~\eqref{ne} is
equivalent to SMCT without memory functions
when we omit the terms which include the self-energies.
However, this equation is
quite complicated, because the self-energies are
included in $F_{\htt\rho}$, $F_{\hg_\a\rho}$ and $F_{\hn_\a\rho}$.
Therefore, we restrict our interest to
the calculation of the self-energies in the
first-loop order perturbation in the latter part of this paper.

In the first-loop order perturbation,
the self-energy $\Sigma_{\hat\phi_1 \hat\phi_1'}$
is expressed as
\begin{eqnarray}
 \label{sig_psi1}
  \Sigma_{\hat\phi_1 \hat\phi_1'} \left(X_1,X_{1}' \right)
  &=&
  \frac{1}{2} \sum_{\phi_2,\phi_3,\phi_{2}',\phi_{3}'}
  \int dX_2 dX_3 dX_{2}' dX_{3}' \,
  \nonumber \\ && 
  V_{\hat\phi_1 \phi_2 \phi_3} \left(X_1,X_2,X_3 \right)
  V_{\hat\phi_{1}' \phi_{2}' \phi_{3}'}
 \left(X_{1}',X_{2}',X_{3}'\right)
G_{\phi_2 \phi_{2}'} \left(X_2-X_{2}'\right) 
G_{\phi_3 \phi_{3}'} \left(X_3-X_{3}'\right),
\end{eqnarray}
where $\hat\phi_1$ or $\hat\phi_1'$ is the one of the components of
$\hat{\vphi}$, and $\phi_i$ or $\phi'_i$
is the one of the components of $\vphi$.
Here 
the three-point vertex $V_{\hat\phi_1 \phi_2 \phi_3}$ is defined by 
\begin{equation}
 \label{threepv}
 V_{\hat\phi_1 \phi_2 \phi_3} \left(X_1,X_2,X_3\right)
  \equiv
  \frac{\delta^3 {S}_{ng} [\vphi]}{\d \phi_1 \left(X_1\right)
  \d \phi_2\left(X_2\right) \d \phi_3\left(X_3\right)}.
\end{equation}
We list the all three-point vertices
which include $\hat{\vphi}$ in APPENDIX~\ref{tpv}.
Note that there are no four-point correlation functions 
including both of $\hat\phi_1$ and $\hat\phi_1'$.

\begin{figure}[t]
 \centering
 \includegraphics[width=80mm]{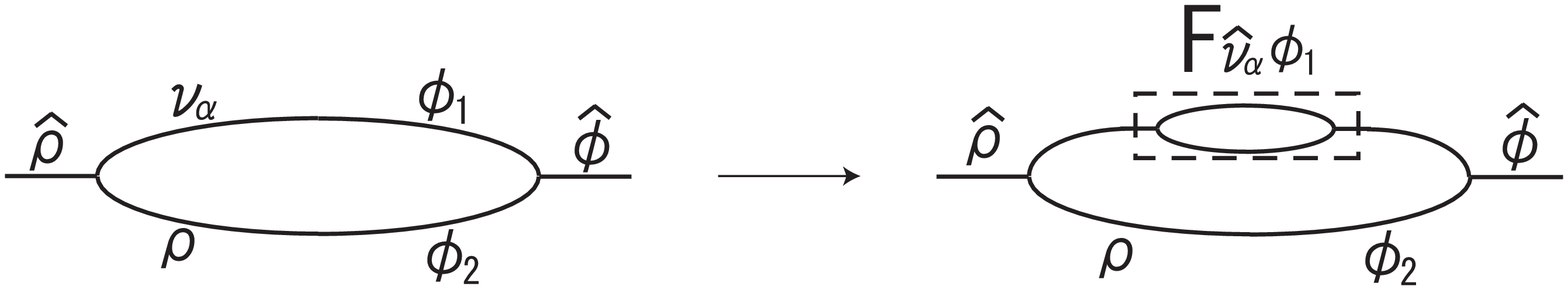}
 \caption{One of the diagrams of
 the self-energy $\Sigma_{\hr \hat\phi}$ which is produced by
 the vertex $V_{\hr \r \nu_\a}$.
 From the eq.~\eqref{kiso4'},
 the first-loop order self-energy
 in the left-hand side of the figure can be treated as
 second-loop order as figured in the right-hand side of the figure.
 }
 \label{fig1}
\end{figure}

Within the first-loop order approximation,  
$F_{\hat\phi\phi}$ in eq.~\eqref{fff} is reduced to
\begin{eqnarray}
 F_{\hat\phi\phi} \left(\vk,t\right)
  &\simeq&
  \int^t_0 ds \, \biggl\{
  \Sigma_{\hp\htt}\left(\vk,t-s\right) \dd_s G_{\r\phi}
  \left(\vk,s\right)
  - (T\r_0)^{-1} \Sigma_{\hp\hg_\a}\left(\vk,t-s\right)
  G_{g_\a\phi}\left(\vk,s\right) \nonumber \\
 && - \Sigma_{\hp\hg_\a}\left(\vk,t-s\right)
  F_{\hn_\a \phi}\left(\vk,s\right)
  + \Sigma_{\hp\hn_\a}(\vk,t-s)
  \left\{
   \dd_s G_{g_\a \phi}(\vk,s)
   + ik_\a T\r_0
   (K(\vk)G_{\r\phi}(\vk,s)+G_{\t\phi}(\vk,s))
  \right\}
 \biggr\} \nonumber \\
 &\simeq& \int^t_0 ds \, 
  \biggl\{
  \Sigma_{\hp\htt}\left(\vk,t-s\right) \dd_s G_{\r\phi}
  \left(\vk,s\right) 
- (T\r_0)^{-1} \Sigma_{\hp\hg_\a}\left(\vk,t-s\right)
  G_{g_\a\phi}\left(\vk,s\right) 
\nonumber \\ &&
 - \Sigma_{\hp\hn_\a} (\vk,t-s)
  \r_0^{-1} L_{\a\b}  G_{g_\b \phi} (\vk,s)
  \biggr
  \}.
  \label{fff''}
\end{eqnarray}
We have used \eqref{kiso3}, \eqref{kiso2'} and \eqref{kiso4'}
and eliminate higher order contributions
 to obtain the final expression.
We have also used the \eqref{kiso4'} and
the relation
\begin{equation}
 \Sigma_{\hp\hr}(\vk,t) \simeq
  - ik_\a T\r_0 \Sigma_{\hp\hn_\a}(\vk,t),
\label{identity2}
\end{equation}
which is only valid in the first-loop order, for the first equality in eq.~\eqref{fff''}.
Let us show the expression \eqref{identity2}.
The self-energy $\Sigma_{\hp\hr}$ should contain the vertex $V_{\hr\r g_\a}$ or $V_{\hr\r \nu_\a}$.
Among these two vertices, 
the vertex $V_{\hr\r \nu_\a}$ is irrelevant within
the first-loop order calculation. Indeed, the self-energy with the
vertex $V_{\hr\r \nu_\a}$ should contain
the propagator $G_{\nu_\a\phi}$ which is equal to $F_{\nu_\a\phi}$ from eq.~\eqref{kiso4'}
and is the first loop order.
Thus, eventually, the self-energy becomes higher order correction as
 in Fig.~\ref{fig1}.
On the other hand, from eqs. \eqref{D1} and \eqref{D9}, the vertices $V_{\hr\r g_\b}$ and
$V_{\hn_\a \r g_\b}$ satisfy the relation
\begin{equation}
 V_{\hr\r g_\b} (X_1,X_2,X_3)
  = T\r_0 \nabla_{\vr_1 \a} V_{\hn_\a \r g_\b} (X_1,X_2,X_3).
\end{equation}
Thus, we reach the relation~\eqref{identity2}.

Let us calculate some typical terms, such as
$F_{\htt\r} \left(\vk,t\right)$, which appear on
the RHS of eq.~\eqref{ne} in the long time behavior under
the first-loop order approximation.
For simplicity, we assume that the correlations include the momentum
decay fast enough to be negligible in the long time behavior.
For this purpose, at first, we calculate $\Sigma_{\htt\htt}\left(\vk,t\right)$.
Among the three-point vertex functions listed in APPENDIX~\ref{tpv},
there are only two vertices~\eqref{vhtt1} and~\eqref{vhtt2}, which
include $\htt$. 
Substituting (\ref{vhtt1}) and (\ref{vhtt2}) into (\ref{sig_psi1}) with
$\hat{\phi}_1=\hat{\phi}_{1}'=\hat\theta$, the expression of
$\Sigma_{\htt\htt}\left(\vk,t\right)$ at the first-loop order is given by
\begin{eqnarray}
 \Sigma_{\htt\htt}\left(\vk,t\right)
  &=& \!
  \int \frac{d\vq}{\left(2\pi\right)^3} \,
  \Biggl\{
  \frac{1}{2 m^2 \r_0^4} G_{\r\r}\left(\vq,t\right)
  G_{\r\r}\left(\vk-\vq,t\right)
  \nonumber \\ && \!
  + \frac{1}{T m \r_0^4}
  \left(
   G_{g_\a\r}\left(\vq,t\right)G_{g_\a\r}\left(\vk-\vq,t\right)
   + G_{\r g_\a}\left(\vq,t\right)G_{\r g_\a}\left(\vk-\vq,t\right) \right)
  \nonumber \\ && \!
  + \frac{2}{T^2 \r_0^4} G_{g_\a g_\b}\left(\vq,t\right)
  G_{g_\a g_\b}\left(\vk-\vq,t\right) \Biggr\}.
\label{eq48}
\end{eqnarray}

In the limit $t \to \infty$, thus,
the first term of $F_{\htt\r}\left(\vk,t\right)$
in eq.~\eqref{fff''} can be approximated by
\begin{eqnarray}
 \int^t_0 ds \,
  \Sigma_{\htt\htt}\left(\vk,t-s\right) \dd_s G_{\r\r}\left(\vk,s\right)
 &\simeq&
  \Sigma_{\htt\htt}\left(\vk,t\right) \int^t_0 ds \,
  \dd_s G_{\r\r}\left(\vk,s\right)
  \nonumber \\ 
  &\simeq&
   \frac{G_{\r\r}\left(\vk,t\right) - G_{\r\r}\left(\vk,0\right)}{2 m^2 \r_0^4}
  \int \frac{d\vq}{\left(2\pi\right)^3} \,
  G_{\r\r}\left(\vq,t\right)G_{\r\r}\left(\vk-\vq,t\right).
  \label{fterm}
\end{eqnarray}
Here, the last expression is obtained from the
assumption that the correlations include momentum
decay fast enough.

Similarly, with the aid of
(\ref{sig_psi1}) and (\ref{vhtt1})-(\ref{vhglast}), 
$\Sigma_{\htt\hg_\a}\left(\vk,t\right)$ 
at the first-loop order calculation 
reduces to
\begin{eqnarray}\label{eq52}
  \Sigma_{\htt\hg_\a}\left(\vk,t\right)
 &\simeq&  \!
 - \int \frac{d\vq}{\left(2\pi\right)^3} \,
 \frac{iT}{m\r_0^2} q_\a \Bigl(
                           K\left(\vq\right) G_{\r\r}\left(\vq,t\right)
                           + G_{\t\r}\left(\vq,t\right) \Bigr)
  G_{\r\r}\left(\vk-\vq,t\right) \nonumber \\
 &=&
   - \frac{iT}{m\r_0^2} \int 
   \frac{d\vq}{\left(2\pi\right)^3} \, q_\a 
  \Bigl(
   K \left(\vq \right)
   G_{\r\r}\left(\vq,t\right)
   +
   F_{\htt\r}\left(\vq,t\right) \Bigr)
  G_{\r\r}\left(\vk-\vq,t\right)
  \nonumber \\
 &\simeq&
  - \frac{iT}{m\r_0^2} \int \frac{d\vq}{\left(2\pi\right)^3} \,
  q_\a
  K \left(\vq \right) G_{\r\r}\left(\vq,t\right)
  G_{\r\r}\left(\vk-\vq,t\right) \nonumber \\
 &=&
  - \frac{iT k_{\beta}}{mk^2\r_0^2} \int \frac{d\vq}{\left(2\pi\right)^3} \,
  k_{\beta} q_\a
  K \left(\vq \right) G_{\r\r}\left(\vq,t\right)
  G_{\r\r}\left(\vk-\vq,t\right),
  \label{shtthg}
\end{eqnarray}
in the limit $t \to \infty$.
The first equality in (\ref{eq52})  comes from the assumption that
the correlations include momentum decay fast enough.
For the second equality in (\ref{eq52}) we have used 
eq.~(\ref{kiso2'}).
To obtain the third equality in (\ref{eq52}) 
we have ignored the contribution from $F_{\htt\r}$.
This simplification can be justified at the first-loop order
approximation,  because
$F_{\htt\r}$ is the first or the above loop order function.
To obtain the last expression in \eqref{eq52}, we have used the fact 
that the density correlation function depends on time and the absolute
value of the wave vector.
From eq.~\eqref{shtthg}, the second term of $F_{\htt\r}\left(\vk,t\right)$
in eq.~\eqref{fff''} becomes
\begin{eqnarray}
  && - \frac{1}{T\r_0}
  \int^t_0 ds \, \Sigma_{\htt\hg_\a}\left(\vk,t-s\right)
  G_{g_\a\r}\left(\vk,s\right)
  \nonumber \\
  &\simeq&
 \frac{i}{m k^2 \r_0^3} \int \frac{d\vq}{\left(2\pi\right)^3} \,
  k_\a q_\a
  K \left(\vq \right) G_{\r\r}\left(\vq,t\right)
  G_{\r\r}\left(\vk-\vq,t\right)
 \int^t_0 ds \, k_\b G_{g_\b\r}\left(\vk,s\right)
  \nonumber \\
  &=&
  \frac{-1}{m k^2 \r_0^3} \int \frac{d\vq}{\left(2\pi\right)^3} \, k_\a q_\a
  K \left(\vq \right) G_{\r\r}\left(\vq,t\right)
  G_{\r\r}\left(\vk-\vq,t\right)
  \int^t_0 ds \, \dd_s G_{\r\r}\left(\vk,s\right)
  \nonumber \\ 
  &=&
   - \frac{G_{\r\r}\left(\vk,t\right) - G_{\r\r}\left(\vk,0\right)}{m k^2 \r_0^3}
  \int \frac{d\vq}{\left(2\pi\right)^3} \, k_\a q_\a
  K \left(\vq\right) G_{\r\r}\left(\vq,t\right)
  G_{\r\r}\left(\vk-\vq,t\right), \label{sterm}
\end{eqnarray}
where we have used
eq.~\eqref{kiso1'} in the second equality.
The last term of $F_{\htt\r}$ in eq.~\eqref{fff''}
is zero because
$\Sigma_{\htt\hn_\a}(\vk,t)$ and $ G_{g_\b \r}(\vk,t)$ are
zero in the long time limit.

Thus, we obtain the expression for $F_{\htt\r}\left(\vk,t\right)$
in eq.~\eqref{fff''} from the eqs.~\eqref{fterm} and \eqref{sterm}
at the first-loop order as
\begin{eqnarray}
 F_{\htt\r}\left(\vk,t\right)
  &=& 
  \int \frac{d\vq}{\left(2\pi\right)^3} \,
  \left\{
   \frac{1}{2m^2\r_0^4}
  - \frac{1}{mk^2\r_0^3} k_\a q_\a K \left(\vq\right)
  \right\}
  G_{\r\r}\left(\vq,t\right) G_{\r\r}\left(\vk-\vq,t\right)
  \left(G_{\r\r}\left(\vk,t\right) - G_{\r\r}\left(\vk,0\right)\right),
   \label{firsthttr}
\end{eqnarray}
in the limit $t \to \infty$. 
Similarly, we evaluate $i k_\a F_{\hg_\a \r}\left(\vk,t\right)$
and $F_{\hn_\a \r}\left(\vk,t\right)$
within the first-loop order as
\begin{eqnarray}
  i k_\a F_{\hg_\a \r}\left(\vk,t\right)
  &=& 
  \int \frac{d\vq}{\left(2\pi\right)^3}  \,
  \Bigl\{
  - \frac{T}{m\r_0^2} k_\a q_\a K \left(\vq\right)
  + \frac{T}{2k^2 \r_0}
  \left(k_\a q_\a K \left(\vq\right)
   +
   k_\a \left(k_\a-
	 q_\a\right)
   K \left(\vk-\vq \right)\right)^2
  \Bigr\}
   \nonumber \\  &&
   \times G_{\r\r}\left(\vq,t\right) G_{\r\r}\left(\vk-\vq,t\right)   \left(G_{\r\r}\left(\vk,t\right) - G_{\r\r}\left(\vk,0\right)\right),
  \label{firsthgr}
\end{eqnarray}
and
\begin{equation}
 F_{\hn_\a \r}\left(\vk,t\right) = 0,
\end{equation}
in the limit $t \to \infty$.

From the above expressions for
$F_{\htt\r}\left(\vk,t\right)$,
$F_{\hg_\a\r}\left(\vk,t\right)$ and
$F_{\hn_\a\r}\left(\vk,t\right)$,
we can evaluate RHS of eq.~\eqref{ne} in the limit  $t\to \infty$.
We note that the first and the second terms on the LHS of
eq.~\eqref{ne} are zero in the long time limit,
because 
the time derivatives of the density correlation functions
should be zero in such a region.
Therefore, the self-consistent
equation of the nonergodic parameter $f(\vk)$, which is defined by
\begin{equation}
 \label{defnp}
 f \left(\vk\right)
 \equiv \lim_{t\to\infty} \frac{G_{\r\r}\left(\vk,t\right)}{G_{\r\r}\left(\vk,0\right)},
\end{equation}
is obtained as
\begin{equation}
 \label{sce}
 f \left(\vk\right) = \frac{M\left(\vk\right)}{1+M\left(\vk\right)},
\end{equation}
where
\begin{equation}
 M\left(\vk\right) \equiv
  \frac{\r_0 S\left(\vk\right)}{2 m k^4} \int \frac{d \vq}{\left(2\pi\right)^3} \,
  {V}_{\vk,\vq}^2 S\left(\vp\right)S\left(\vq\right) f
  \left(\vp\right) f\left(\vq\right) ,
\end{equation}
with
\begin{equation}\label{v_k,q}
 {V}_{\vk,\vq} \equiv k_\a \left(q_\a c\left(\vq\right) + p_\a c\left(\vp\right)\right),
\end{equation}
where we have used $\vp \equiv \vk-\vq$ and the static structure factor (\ref{ssf}). 
This set of self-consistent equations~\eqref{defnp}-(\ref{v_k,q}) for the  nonergodic parameter is
equivalent to that in SMCT.

\section{Discussion and Conclusion}\label{sDIS}

\subsection{Discussion}

In this paper, we formulate the FDR-preserving field theory for FNH.
The obtaining SMCT under the first-loop order approximation in the previous section, is a successful first step
to construct a correct theory beyond SMCT. However, we still have some unclear points in the analysis. Let us
discuss such unclear points through the comparison with other field theoretical approaches. 

To analyze the SD equation,
we employ the important assumption that
the correlations include the momentum are negligibly small in  the long time region.
This assumption is crucial to discuss whether ENE transition exists. 
Indeed, if we assume that the contributions from momentum are negligible,
FNH equations are reduced to the Dean-Kawasaki equation,
as indicated by Kawasaki~\cite{kawasaki1994}.
On the other hand,
there are several indirect evidences for the justification of this approximation.
First, we note that a numerical simulation exhibits fast relaxations of  
the correlations including the momentum~\cite{lust1993}.
Second, it is known that the density-density correlation $G_{\r\r}$ is connected only to the correlation in the longitudinal mode 
$k_\a G_{g_\a \phi}$, which is proportional to $\dd_t G_{\r \phi}$.
Since the system is almost stationary, any terms including the time derivative are small. 
Thus, the contributions from the correlation including
the momentum can be
ignored in the slow dynamics in the motion of the density field.

\begin{table*}[t]
 \label{comparison}
  \begin{center}
   \begin{ruledtabular}
   \begin{tabular}{p{16mm}p{7mm}p{31mm}p{10mm}p{20mm}p{10mm}p{28mm}
    p{10mm}p{25mm}} 
    paper & & (i) model & & (ii) FDR & & (iii) approximations & &
    (iv)nonergodic parameter\\
    \hline
    DM~\cite{dm} & & FNH & & non-preserve & &
                            non-perturbative & & 0 \\
   \hline
    SDD~\cite{sdd} & &
            FNH with violation of Galilean invariance & &
                    preserve  & & first-loop order,
                            $G_{g \phi} \to 0$ & &
                                    0 \\
   \hline
    ABL~\cite{abl} & & Dean-Kawasaki and FNH & & preserve
                    & & first-loop order, $G_{g \phi} \to 0$ & & 1\\ 
    \hline
    KK~\cite{kk} & & Dean-Kawasaki & & preserve & & first-loop order & &
                                    $f_{SMCT}(k)$\\
   \hline
   This work & & FNH & & preserve & & first-loop order, $G_{g \phi} \to 0$
               & & $f_{SMCT}(k)$ \\
   \end{tabular}
    \end{ruledtabular}
  \end{center}
 \caption{The comparison of our results
 with Das and Mazenko (DM)~\cite{dm},
 SDD~\cite{sdd} and ABL~\cite{abl} and Kim and Kawasaki (KK)~\cite{kk}.
We list the results on the four points: 
 (i)  the used model,
 (ii) whether FDR is preserved in the perturbation,
 (iii) the used approximation to derive the
 density correlation function in the limit $t \to \infty$, and 
(iv) the behavior of the nonergodic parameter $f(k)$. 
 Here, the expression $G_{g\phi} \to 0$ means that
 the correlations include the momentum becomes zero in the long time limit.
  $f_{SMCT} (k)$ means the nonergodic parameter is equivalent to that of
 SMCT.
}
\end{table*}

Let us compare  our results with those obtained in  other
field theoretical researches, taking into account following four aspects.
(i) Is the basic model adequate?
(ii) Is  the analysis FDR preserved?
(iii) Are the approximations to analyze the
density correlation function valid in the limit $t \to \infty$?
(iv) What is the behavior of the
density correlation function in the limit $t \to \infty$?
The results of the comparison are summarized in Table I.

First, we compare ours with ABL~\cite{abl}.
As can be seen in Table I,
they predicted that the nonergodic
parameter is unity, which is independent of the wave number.
This result is clearly in contrast to  the observations
in experiments and simulations.
This result might be caused by the definition
of the new set of additional fields.
Indeed, we have introduced
$\theta$ in eq.~\eqref{myt} and
$\vn$ in eq.~\eqref{mynu}, while ABL~\cite{abl} 
\begin{equation}\label{abl-49}
 \theta_{ABL} \equiv  \frac{\d F}{\d \r}, 
  \quad {\rm{and}} \quad
  \nu_{\alpha,ABL} \equiv \frac{\partial F}{\partial g_{\alpha}}.
\end{equation}
These new variables include the linear terms of
$\d\r$ and $\vg$, but
our additional variables $\t$ and $\vn$
do no include the linear terms.
As a result, the order of the correlations, which include
the new variables $\t$ and $\vn$, differ from ours.
Indeed, $G_{\t\phi}$ and $G_{\nu_\a \phi}$ are the first or
the above loop order in eq.~\eqref{kiso2'} and eq.~\eqref{kiso4'}, 
while $G_{\t\phi}$ and $G_{\nu_\a \phi}$
include the tree diagrams in the calculation of ABL.
Therefore, we suggest that eq.~\eqref{abl-49} is not appropriate, 
but we should use eqs.~\eqref{myt} and \eqref{mynu}.

Next, we compare our results with 
those by Kim and Kawasaki~\cite{kk}.
Their method is almost parallel to that we have used here.
However, their basic equation is not FNH equations
but Dean-Kawasaki equation.
Thus, their MCT equation without interactions is the diffusion equation.
On the other hand, our MCT equation~\eqref{ne} without memory kernel
is equation for a damped oscillator. 
Thus, the existence of the momentum conservation equation 
in the basic equation
naturally leads to the existence of the acceleration term
in MCT equation.

Third, let us compare our results with those by SDD~\cite{sdd},
in which they used a simplified model of FNH.
Although the approximation used here is similar to that used by SDD,
the Galilean invariance is not preserved in their model equation.
It implies that the violations of these conservation laws
cause the artificial cutoff mechanism.

Fourth, we compare ours with Das and Mazenko~\cite{dm}.
One of important differences
between ours and theirs is that they regard $\vv{V}$,
which satisfies the constraint $\vv{V} \equiv \vv{g}/\rho$,
as one collective variable.
On the other hand, we introduce the new field variables $\theta$ in
eq.~\eqref{myt} and $\vn$ in eq.~\eqref{mynu} to satisfy FDR.
Thus, their explicit expressions differ from ours.
Second, let us discuss about their conclusion on the existence of
the cutoff mechanism, {\it i.e.}
$G_{\r\r} \propto G_{\r\hr} = 0$ in the long time limit.
As was indicated by ABL~\cite{abl},  the relation
$G_{\r\r} \propto G_{\r\hr}$, 
used by Das and Mazenko (see eq.~(6.62) in \cite{dm}),
is not preserve FDR.
Thus, we cannot conclude $G_{\r\r}=0$ from the relation
$G_{\r\hr}=0$.
However, the relation $G_{\r\hr}=0$ might be valid, which
was derived from their non-perturbative analysis.
On the other hand, from eq.~\eqref{relg1},
our FDR-preserving calculation under the first-loop order perturbation 
suggests $G_{\r\hr} = K(\vv{k}) G_{\r\r} + G_{\t\r} \ne 0$,
where we have used the numerical result of the non-ergodic parameter.
Thus, our result in the first-loop order perturbation on
$G_{\r\hr}$ is not consistent with Das and Mazenko. 
To resolve this discrepancy between their theory~\cite{dm} and ours, or
to verify  their analysis on the cutoff mechanism, 
we need to find some identical relations without using the approximations.

\subsection{Conclusion}

In this paper, we reformulate a FDR-preserving
field theory starting from FNH.
When we assume that the correlations include the momentum decay fast enough,
we have shown that there exists ENE transition and the nonergodic parameter
is equivalent to SMCT under the first-loop order approximation.
These results give the theoretical basis of SMCT,
and provides a route to go beyond SMCT. 
If we analyze correlations in  higher-loop orders 
we believe that we will be able to construct a correct theory 
to explain the experimental and numerical results.

\begin{acknowledgments}
The authors would like to express sincere gratitude to
B. Kim, K. Kawasaki, K. Miyazaki, S. P. Das, H. Wada, H. Ueda, T. Ohkuma,
T. Nakamura and T. Kuroiwa for fruitful discussions and useful comments.
This work is partially supported by the Ministry of Education,
 Culture, Sports, Science and Technology (MEXT) of Japan (Grant
 No. 18540371)
 and by Grant-in-Aid for the global COE program
``The Next Generation of Physics, Spun from Universality and Emergence''
 from MEXT of Japan.
\end{acknowledgments}

\appendix

\section{Time reversal transformation and FDR}\label{mtr}

From the linearity of the time reversal transformation,
we can rewrite eq.~\eqref{reversal2} 
\begin{equation}
 \label{trmatrix}
 \begin{pmatrix}
  \rho \\
  \hat\rho \\
  \theta \\
  \hat\theta\\
  g_\a \\
  \hat{g}_\a \\
  \nu_\a \\
  \hat\nu_\a
 \end{pmatrix}
 \to
 \begin{pmatrix}
   1&0&0&0&0&0&0&0 \\
   K&-1&1&0&0&0&0&0 \\
   0&0&1&0&0&0&0&0 \\
   \dd_t&0&0&1&0&0&0&0 \\
   0&0&0&0&-\delta_{\a\b}&0&0&0 \\
   0&0&0&0&-\frac{1}{T\rho_0} \delta_{\a\b}&\delta_{\a\b}&-\delta_{\a\b}&0 \\
   0&0&0&0&0&0&-\delta_{\a\b}&0 \\
   0&0&0&0&-\delta_{\a\b}\dd_t &0&0&-\delta_{\a\b} \\
 \end{pmatrix}
 \begin{pmatrix}
   \rho \\
   \hat\rho\\
   \theta\\
   \hat\theta\\
   g_\b \\
   \hat{g}_\b \\
   \nu_\b \\
   \hat\nu_\b
 \end{pmatrix}
 .
\end{equation}
When we express this time reversal transformation matrix as $\mat{O}$,
eq.~\eqref{reversal2} can be represented by $\vpsi \to \mat{O} \, \vpsi$.
It implies that the correlation functions satisfies
\begin{equation}
 \label{trg}
 \mat{G} \left(-t\right) = \mat{O} \, \mat{G}\left(t\right) \, \mat{O}^T.
\end{equation}
Similarly, the self-energy satisfies
\begin{equation}
 \label{trs}
  \mat{\Sigma} \left(-t\right)
  = \left(\mat{O}^T\right)^{-1} \, \mat{\Sigma}(t) \, \mat{O}^{-1},
\end{equation}
where we have used the relation ~\eqref{sd} or
\begin{equation}
 \mat{G}^{-1} = \mat{G}_0^{-1} - \mat{\Sigma}.
\end{equation}

Here, we avoid to write all the components of time reversal
symmetry relations
\eqref{trg} and \eqref{trs}, because they are long and tedious equations.
Instead, we present some typical relations, which are
necessary for the calculation of $\htt\r$ component of the SD equation;
\begin{equation}
 \label{relg1}
 G_{\hr\r}\left(\vk,t\right) = \Theta\left(-t\right)\left(K\left(\vk\right) G_{\r\r}\left(\vk,t\right) + 
G_{\t\r}\left(\vk,t\right)\right),
\end{equation}
\begin{equation}
 G_{\htt\r}\left(\vk,t\right) = - \Theta\left(-t\right) \dd_t G_{\r\r}\left(\vk,t\right),
\end{equation}
\begin{equation}
 G_{\hg_\a \r}\left(\vk,t\right) = \Theta\left(-t\right)
\left(\frac{1}{T\r_0} G_{g_\a\r}\left(\vk,t\right)+ G_{\hn_\a \r}\left(\vk,t\right)\right),
\end{equation}
\begin{equation}
 \label{relg4}
 G_{\hn_\a\r}\left(\vk,t\right) = - \Theta\left(-t\right) \dd_t G_{g_\a\r}\left(\vk,t\right).
\end{equation}
Here, we summarize some relevant relations among the self-energies
\begin{equation}
 \label{rels1}
 \Sigma_{\hat\theta\rho}\left(\vk,t\right) =
  \Theta\left(t\right)
  \left(\dd_t \Sigma_{\hat\theta\hat\theta}\left(\vk,t\right)
  - K\left(\vk\right) \Sigma_{\htt\hr}\left(\vk,t\right)\right),
\end{equation}
\begin{equation}
 \Sigma_{\hat\theta\theta}\left(\vk,t\right) =
  - \Theta\left(t\right) \Sigma_{\htt\hr}\left(\vk,t\right),
\end{equation}
\begin{equation}
 \Sigma_{\hat\theta {g}_\a} \left(\vk,t\right) =
  - \Theta\left(t\right) \left(\frac{1}{T\rho_0}
           \Sigma_{\hat\theta\hat{g}_\a}\left(\vk,t\right) -
  \dd_t \Sigma_{\hat\theta\hat\nu_\a}\left(\vk,t\right)\right),
\end{equation}
\begin{equation}
 \label{rels4}
 \Sigma_{\hat\theta\nu_\a} \left(\vk,t\right) =
  -\Theta\left(t\right) \Sigma_{\hat\theta \hat{g}_\a} \left(\vk,t\right),
\end{equation}
\begin{equation}
 \label{relt1}
 \Sigma_{\htt\htt}\left(\vk,t\right) = \Sigma_{\htt\htt}\left(\vk,-t\right),
\end{equation}
\begin{equation}
 \label{relt2}
 \Sigma_{\htt\hn_\a}\left(\vk,t\right) = - \Sigma_{\htt\hn_\a}\left(\vk,-t\right).
\end{equation}

\section{Calculation of the  FDR-preserving SD equation}\label{csd}

In this APPENDIX, we calculate $\htt\r$ component of
SD equation with the aid of APPENDIX~\ref{mtr}.
From eqs.~\eqref{sg} and (\ref{S_g}) ,
there is only $-\htt\t$ term in $S_g$ which includes  $\htt$.
Therefore, we obtain $(G_0^{-1})_{\htt\phi} (X_1-X_2)= \delta_{\phi\theta}
\delta(X_1-X_2)$.
Thus, $\htt\r$ component of
$\mat{G_0}^{-1} \cdot \mat{G}$ satisfies
\begin{equation}
 \label{httr0}
 [\mat{G_0}^{-1} \cdot \mat{G}]_{\htt\r} \left(\vk,t\right)
 = G_{\t\r}\left(\vk,t\right).
\end{equation}

On the other hand, $\htt\r$ component of $\mat{\Sigma} \cdot \mat{G}$
is expressed as
\begin{eqnarray}
 [\mat{\Sigma} \cdot \mat{G}]_{\htt\r} \left(\vk,t\right)
 &=&
  \int^\infty_{-\infty} ds \,
  \biggl\{\Sigma_{\htt\rho}\left(\vk,t-s\right) G_{\rho\rho}\left(\vk,s\right)
  + \Sigma_{\htt\theta}\left(\vk,t-s\right) G_{\theta\rho}\left(\vk,s\right) 
  \nonumber \\ &&
   + \Sigma_{\htt {g}_\a}\left(\vk,t-s\right) G_{{g}_\a\rho}\left(\vk,s\right)
  +  \Sigma_{\htt\nu_\a}\left(\vk,t-s\right) G_{\nu_\a\rho}\left(\vk,s\right)
   \nonumber \\ &&
 + \Sigma_{\htt\hr}\left(\vk,t-s\right) G_{\hat\r\rho}\left(\vk,s\right)
 + \Sigma_{\htt\hat\theta}\left(\vk,t-s\right) G_{\hat\theta\rho}\left(\vk,s\right)
 \nonumber \\ &&
 + \Sigma_{\htt\hat{g}_\a}\left(\vk,t-s\right) G_{\hat{g}_\a\rho}\left(\vk,s\right)
 +  \Sigma_{\htt\hat\nu_\a}\left(\vk,t-s\right) G_{\hat\nu_\a\rho}\left(\vk,s\right) \biggr\}. \label{httr1}
\end{eqnarray}

By using the relations \eqref{rels1}-\eqref{rels4},  the first four
terms in the right-hand side of eq.~\eqref{httr1} are rewritten as
\begin{eqnarray}
 &&
  - \int^t_{-\infty} ds \,
 \biggl\{
 \left(
 \dd_s \Sigma_{\hat\theta\hat\theta}
 \left(\vk,t-s\right)
  +
  K\left(\vk\right)
  \Sigma_{\htt\hr}\left(\vk,t-s\right) 
\right)
 G_{\rho\rho}\left(\vk,s\right)
   +
   \Sigma_{\htt\hr}\left(\vk,t-s\right) 
G_{\theta\rho}\left(\vk,s\right)
  \nonumber \\ &&
  + \left(\frac{1}{T\rho_0}
		    \Sigma_{\hat\theta\hat{g}_\a}\left(\vk,t-s\right)
     +
  \dd_s \Sigma_{\hat\theta\hat\nu_\a}\left(\vk,t-s\right)
\right)  G_{{g}_\a\rho}
\left(\vk,s\right)
 +
 \Sigma_{\htt\hg_\a}\left(\vk,t-s\right)
 G_{\nu_\a\rho}\left(\vk,s
\right) \biggr\}
 \nonumber \\ &=&
 - \Sigma_{\htt\htt}\left(\vk,0\right) G_{\r\r}\left(\vk,t\right)
 \nonumber \\ &&
 - \int^t_{-\infty} ds \,
 \biggl\{ \Sigma_{\htt\hr}
 \left(\vk,t-s\right)
 \left(K
                 \left(\vk\right) 
G_{\rho\rho}\left(\vk,s\right)
  +
  G_{\theta\rho}\left(\vk,s\right)
\right)
 -
\Sigma_{\hat\theta\hat\theta}\left(\vk,t-s\right) \dd_s 
G_{\rho\rho}\left(\vk,s\right)
 \nonumber \\ &&
+ \Sigma_{\htt\hg_\a}
\left(\vk,t-s\right)
 \left(\frac{1}{T\rho_0}
		 G_{{g}_\a\rho}\left(\vk,s\right)
  +
  G_{\nu_\a\rho}
  \left(\vk,s\right)\right)
 -
 \Sigma_{\hat\theta\hat\nu_\a}
\left(\vk,t-s\right)
 \dd_s G_{{g}_\a\rho}\left(\vk,s\right)
 \biggr\},
\end{eqnarray}
where the boundary terms 
vanish except for
$\Sigma_{\htt\htt}\left(\vk,0\right) G_{\r\r}\left(\vk,t\right)$ because of
$\Sigma_{\htt\hn_\a}\left(\vk,0\right) = 0$ from \eqref{relt2}
and $G_{\r\r}\left(\vk,-\infty\right) =G_{g_\a \r}\left(\vk,-\infty\right) = 0$.

Similarly, from the relations \eqref{relg1}-\eqref{relg4} the last four terms on right-hand side
of eq.~\eqref{httr1} become
\begin{eqnarray}
  && 
\int^0_{-\infty} ds \,
  \biggl\{ 
\Sigma_{\htt\hr}(\vk,t-s) 
\left(
K(\vk) G_{\r\r}(\vk,s) + G_{\theta\r}(\vk,s)
\right)
 -\Sigma_{\hat\theta\hat\theta}(\vk,t-s) 
\dd_s G_{\r\r}(\vk,s)
 \nonumber \\ 
&& + \Sigma_{\htt\hg_\a}\left(\vk,t-s\right)
 \left(
\frac{1}{T\rho_0} G_{{g}_\a\rho}(\vk,s)
 + G_{\nu_\a\rho}(\vk,s)
\right)
 -\Sigma_{\hat\theta\hat\nu_\a}\left(\vk,t-s\right) 
\dd_s G_{{g}_\a\rho}\left(\vk,s\right) 
\biggr\}.
  \label{httr3}
\end{eqnarray}
 From eqs.~\eqref{httr0}-\eqref{httr3} , we obtain $\htt\r$ component of
the FDR-preserving SD equation as 
\begin{eqnarray}
G_{\t\r}\left(\vk,t\right) +
  \Sigma_{\htt\htt}\left(\vk,0\right) G_{\r\r}\left(\vk,t\right)
   &=& 
   - \int^t_{0} ds \,
   \biggl\{\Sigma_{\hg_\b\hr}\left(\vk,t-s\right) \left(K\left(\vk\right)
			      G_{\rho\rho}\left(\vk,s\right) + G_{\t\r}\left(\vk,s\right)\right)
    \nonumber \\ 
   && 
   - \Sigma_{\hg_\b\htt}\left(\vk,t-s\right) \dd_s G_{\r\r}\left(\vk,s\right)
   - \Sigma_{\hg_\b\hn_\a}\left(\vk,t-s\right) \dd_s G_{g_\a\r}\left(\vk,s\right)
   \nonumber \\ 
   && +
  \Sigma_{\hg_\b\hg_\a}\left(\vk,t-s\right)
  \left(\frac{1}{T\r_0} G_{g_\a\r}\left(\vk,s\right) + G_{\nu_\a\r}\left(\vk,s\right)\right)
   \biggr\}.
   \label{httr}
\end{eqnarray}
Similarly, other components of SD equation
can be obtained with the aid of the time reversal symmetry~\eqref{trmatrix}.

\section{Some exact relations of equal-time correlation functions and self-energies}\label{scf}

In this Appendix,
we derive some relations for the equal-time correlation functions
and the self-energies
from the effective free energy $F$.
Here, we note 
that the mean value 
is calculated by the canonical average over $F$.
Since the equal-time correlation function satisfies
\begin{eqnarray}
 \left\langle \dr(\vr) \frac{\d F}{\d \r(\vr')} \right\rangle
 &=& T \delta(\vr - \vr') \nonumber \\
 &=& T \left\langle \dr(\vr) K * \dr(\vr') \right\rangle
  + T \left\langle \dr(\vr) \t(\vr') \right\rangle.
\end{eqnarray}
The Fourier transform of this equation becomes
\begin{eqnarray}
 \left\langle \dr(\vk) \t(-\vk) \right\rangle = 0,
\end{eqnarray}
where we 
have used 
the relation $K(\vk) = G^{-1}_{\r\r}(\vk,0)$.
Thus, 
we obtain 
\begin{equation}\label{C5}
 G_{\r\t}(\vk,0) = 0.
\end{equation}
Substituting (\ref{C5}) into 
eq.~\eqref{kiso2}, 
and setting $t=0$, we obtain
\begin{equation}\label{C4_old}
 \Sigma_{\htt\htt}(\vk,0) = 0.
\end{equation}

Similarly, 
from 
the relations
\begin{eqnarray}
 \label{gfg}
  \left\langle g_\a(\vr) \rho(\vr')
   \frac{\d F}{\d g_\b(\vr')} \right\rangle 
  &=& T \r_0 \delta_{\a\b} \delta(\vr - \vr') \nonumber\\
 &=& \left\langle g_\a(\vr) g_\b(\vr') \right\rangle ,
\end{eqnarray}
and
\begin{eqnarray}
 \label{gfg2}
  \left\langle g_\a(\vr) \frac{\d F}{\d g_\b(\vr')} \right\rangle 
  &=& T \delta_{\a\b} \delta(\vr - \vr') \nonumber\\
 &=& \r_0^{-1} \left\langle g_\a(\vr) g_\b(\vr') \right\rangle
  + T \left\langle g_\a(\vr) \nu_\b(\vr') \right\rangle,
\end{eqnarray}
we obtain
\begin{equation}\label{C13}
 G_{g_\a \nu_\b} (\vk,0) = G_{\nu_\b g_\a} (\vk,0) = 0.
\end{equation}
Substituting eq.~(\ref{C13}),
into eq.~\eqref{kiso4} at $t = 0$,
we obtain
\begin{equation}\label{C9_old}
 \Sigma_{\hn_\a \hn_\b} (\vk,0) = 0.
\end{equation}

\section{The list of three-point vertices}\label{tpv}

In  this Appendix, we write down the all of the
$\hat{\vphi}$ included
three-point vertices 
which are defined by~\eqref{threepv} 
\begin{eqnarray}
 V_{\hr\r g_\a}\left(X_1,X_2,X_3\right) &=& - \r_0^{-1} 
 \nabla_{\vr_1 \a} \delta\left(X_1-X_2\right) \delta\left(X_1-X_3\right), \label{D1} \\
 V_{\hr\r \nu_\a}\left(X_1,X_2,X_3\right) &=&
  - T \nabla_{\vr_1 \a} \delta\left(X_1-X_2\right) \delta\left(X_1-X_3\right), \label{D2}\\
 V_{\htt\r\r}\left(X_1,X_2,X_3\right) &=& - \frac{1}{m\r_0^2} 
  \delta\left(X_1-X_2\right)
  \delta\left(X_1-X_3\right), \label{vhtt1} \\
 V_{\htt g_\a g_\b}\left(X_1,X_2,X_3\right)
  &=& -
  \frac{1}{T \r_0^2}
  \delta_{\a\b} \delta\left(X_1-X_2\right) \delta\left(X_1-X_3\right),
  \label{vhtt2} \\
  V_{\hg_\a \r \r}\left(X_1,X_2,X_3\right)
   &=& - T
   \biggl\{\delta\left(X_1-X_2\right) \nabla_{\vr_1 \a}
   K\left(X_1-X_3\right) 
   \nonumber \\
 && 
  + \delta\left(X_1-X_3\right) \nabla_{\vr_1 \a}
  K\left(X_1-X_2 \right) \biggr\}, \\
  V_{\hg_\a \r \t}\left(X_1,X_2,X_3\right)
   &=&
   - T \delta\left(X_1-X_2\right) \nabla_{\vr_1 \a}
  \delta\left(X_1-X_3\right), \\
 V_{\hg_\a g_\b g_\c}\left(X_1,X_2,X_3\right)
  &=&
  - \r_0^{-1}
  \biggl\{\delta_{\a\b} \nabla_{\vr_1 \c} 
  [\delta\left(X_1-X_2\right) \delta\left(X_1-X_3\right) ]
  \nonumber \\
 && 
  + \delta_{\b\c} \delta\left(X_1-X_2\right) \nabla_{\vr_1 \a} \delta\left(X_1-X_3\right) 
 \biggr\}, \\
 V_{\hg_\a g_\b \nu_\c}\left(X_1,X_2.X_3\right)
  &=&
  -T \biggl\{\delta_{\a\b} \nabla_{\vr_1 \c}
  [\delta\left(X_1-X_2\right) \delta\left(X_1-X_3\right)]
  \nonumber \\
 && 
  + \delta_{\b\c} \delta\left(X_1-X_2\right) \nabla_{\vr_1 \a} \delta\left(X_1-X_3\right)  \biggr\}, \label{vhglast}\\
 V_{\hn_\a \r g_\b}\left(X_1,X_2,X_3\right) &=&
 -\frac{1}{T \r_0^2} \delta_{\a\b} \delta\left(X_1-X_2\right)
  \delta\left(X_1-X_3\right),\label{D9}
\end{eqnarray}
where $\delta (X_1-X_2) \equiv \delta(\vr_1-\vr_2,t_1-t_2)$ and
$K(X_1-X_2) \equiv \delta(t_1-t_2) K * \delta(\vr_1-\vr_2)$.

\bibliographystyle{apsrevM}
\bibliography{paper}

\ifx\mcitethebibliography\mciteundefinedmacro
\PackageError{apsrevM.bst}{mciteplus.sty has not been loaded}
{This bibstyle requires the use of the mciteplus package.}\fi
\begin{mcitethebibliography}{42}
\expandafter\ifx\csname natexlab\endcsname\relax\def\natexlab#1{#1}\fi
\expandafter\ifx\csname bibnamefont\endcsname\relax
  \def\bibnamefont#1{#1}\fi
\expandafter\ifx\csname bibfnamefont\endcsname\relax
  \def\bibfnamefont#1{#1}\fi
\expandafter\ifx\csname citenamefont\endcsname\relax
  \def\citenamefont#1{#1}\fi
\expandafter\ifx\csname url\endcsname\relax
  \def\url#1{\texttt{#1}}\fi
\expandafter\ifx\csname urlprefix\endcsname\relax\def\urlprefix{URL }\fi
\providecommand{\bibinfo}[2]{#2}
\providecommand{\eprint}[2][]{\url{#2}}

\bibitem[{\citenamefont{Ediger et~al.}(1996)\citenamefont{Ediger, Angell, and
  Nagel}}]{review-ediger}
\bibinfo{author}{\bibfnamefont{M.~D.} \bibnamefont{Ediger}},
  \bibinfo{author}{\bibfnamefont{C.~A.} \bibnamefont{Angell}},
  \bibnamefont{and} \bibinfo{author}{\bibfnamefont{S.~R.} \bibnamefont{Nagel}},
  \bibinfo{journal}{J. Phys. Chem.} \textbf{\bibinfo{volume}{100}},
  \bibinfo{pages}{13200} (\bibinfo{year}{1996})\relax
\mciteBstWouldAddEndPuncttrue
\mciteSetBstMidEndSepPunct{\mcitedefaultmidpunct}
{\mcitedefaultendpunct}{\mcitedefaultseppunct}\relax
\EndOfBibitem
\bibitem[{\citenamefont{Angell et~al.}(2000)\citenamefont{Angell, Ngai,
  McKenna, McMillan, and Martin}}]{review-angell}
\bibinfo{author}{\bibfnamefont{C.~A.} \bibnamefont{Angell}},
  \bibinfo{author}{\bibfnamefont{K.~L.} \bibnamefont{Ngai}},
  \bibinfo{author}{\bibfnamefont{G.~B.} \bibnamefont{McKenna}},
  \bibinfo{author}{\bibfnamefont{P.~F.} \bibnamefont{McMillan}},
  \bibnamefont{and} \bibinfo{author}{\bibfnamefont{S.~W.}
  \bibnamefont{Martin}}, \bibinfo{journal}{J. Appl. Phys.}
  \textbf{\bibinfo{volume}{88}}, \bibinfo{pages}{3113}
  (\bibinfo{year}{2000})\relax
\mciteBstWouldAddEndPuncttrue
\mciteSetBstMidEndSepPunct{\mcitedefaultmidpunct}
{\mcitedefaultendpunct}{\mcitedefaultseppunct}\relax
\EndOfBibitem
\bibitem[{\citenamefont{Debenedetti and Stillinger}(2001)}]{review-debenedetti}
\bibinfo{author}{\bibfnamefont{P.~G.} \bibnamefont{Debenedetti}}
  \bibnamefont{and} \bibinfo{author}{\bibfnamefont{F.~H.}
  \bibnamefont{Stillinger}}, \bibinfo{journal}{Nature}
  \textbf{\bibinfo{volume}{410}}, \bibinfo{pages}{259}
  (\bibinfo{year}{2001})\relax
\mciteBstWouldAddEndPuncttrue
\mciteSetBstMidEndSepPunct{\mcitedefaultmidpunct}
{\mcitedefaultendpunct}{\mcitedefaultseppunct}\relax
\EndOfBibitem
\bibitem[{\citenamefont{G{\"{o}}tze}(1991)}]{review-gotze}
\bibinfo{author}{\bibfnamefont{W.~W.} \bibnamefont{G{\"{o}}tze}}, in
  \emph{\bibinfo{booktitle}{Liquids, Freezing and Glass Transition}}, edited by
  \bibinfo{editor}{\bibfnamefont{J.~P.} \bibnamefont{Hansen}},
  \bibinfo{editor}{\bibfnamefont{D.}~\bibnamefont{Levesque}}, \bibnamefont{and}
  \bibinfo{editor}{\bibfnamefont{J.}~\bibnamefont{Zinn-Justin}}
  (\bibinfo{publisher}{North-Holland}, \bibinfo{address}{Amsterdam},
  \bibinfo{year}{1991})\relax
\mciteBstWouldAddEndPuncttrue
\mciteSetBstMidEndSepPunct{\mcitedefaultmidpunct}
{\mcitedefaultendpunct}{\mcitedefaultseppunct}\relax
\EndOfBibitem
\bibitem[{\citenamefont{Das}(2004)}]{review-das}
\bibinfo{author}{\bibfnamefont{S.~P.} \bibnamefont{Das}},
  \bibinfo{journal}{Rev. Mod. Phys.} \textbf{\bibinfo{volume}{76}},
  \bibinfo{pages}{785} (\bibinfo{year}{2004})\relax
\mciteBstWouldAddEndPuncttrue
\mciteSetBstMidEndSepPunct{\mcitedefaultmidpunct}
{\mcitedefaultendpunct}{\mcitedefaultseppunct}\relax
\EndOfBibitem
\bibitem[{\citenamefont{Reichman and Charbonneau}(2005)}]{review-reichman}
\bibinfo{author}{\bibfnamefont{D.~R.} \bibnamefont{Reichman}} \bibnamefont{and}
  \bibinfo{author}{\bibfnamefont{P.}~\bibnamefont{Charbonneau}},
  \bibinfo{journal}{J. Stat. Mech.} \bibinfo{pages}{P05013}
  (\bibinfo{year}{2005})\relax
\mciteBstWouldAddEndPuncttrue
\mciteSetBstMidEndSepPunct{\mcitedefaultmidpunct}
{\mcitedefaultendpunct}{\mcitedefaultseppunct}\relax
\EndOfBibitem
\bibitem[{\citenamefont{Miyazaki}(2007)}]{review-miyazaki}
\bibinfo{author}{\bibfnamefont{K.}~\bibnamefont{Miyazaki}},
  \bibinfo{journal}{Bussei Kenkyu} \textbf{\bibinfo{volume}{88}},
  \bibinfo{pages}{621} (\bibinfo{year}{2007}), \bibinfo{note}{(in
  Japaneses)}\relax
\mciteBstWouldAddEndPuncttrue
\mciteSetBstMidEndSepPunct{\mcitedefaultmidpunct}
{\mcitedefaultendpunct}{\mcitedefaultseppunct}\relax
\EndOfBibitem
\bibitem[{\citenamefont{Das and Mazenko}(1986)}]{dm}
\bibinfo{author}{\bibfnamefont{S.~P.} \bibnamefont{Das}} \bibnamefont{and}
  \bibinfo{author}{\bibfnamefont{G.~F.} \bibnamefont{Mazenko}},
  \bibinfo{journal}{Phys. Rev. A} \textbf{\bibinfo{volume}{34}},
  \bibinfo{pages}{2265} (\bibinfo{year}{1986})\relax
\mciteBstWouldAddEndPuncttrue
\mciteSetBstMidEndSepPunct{\mcitedefaultmidpunct}
{\mcitedefaultendpunct}{\mcitedefaultseppunct}\relax
\EndOfBibitem
\bibitem[{\citenamefont{G{\"{o}}tze and Sjogren}(1987)}]{gotze1987}
\bibinfo{author}{\bibfnamefont{W.~W.} \bibnamefont{G{\"{o}}tze}}
  \bibnamefont{and} \bibinfo{author}{\bibfnamefont{L.}~\bibnamefont{Sjogren}},
  \bibinfo{journal}{Z. Phys. B: Condens. Matter} \textbf{\bibinfo{volume}{65}},
  \bibinfo{pages}{415} (\bibinfo{year}{1987})\relax
\mciteBstWouldAddEndPuncttrue
\mciteSetBstMidEndSepPunct{\mcitedefaultmidpunct}
{\mcitedefaultendpunct}{\mcitedefaultseppunct}\relax
\EndOfBibitem
\bibitem[{\citenamefont{Schmitz et~al.}(1993)\citenamefont{Schmitz, Dufty, and
  De}}]{sdd}
\bibinfo{author}{\bibfnamefont{R.}~\bibnamefont{Schmitz}},
  \bibinfo{author}{\bibfnamefont{J.~W.} \bibnamefont{Dufty}}, \bibnamefont{and}
  \bibinfo{author}{\bibfnamefont{P.}~\bibnamefont{De}}, \bibinfo{journal}{Phys.
  Rev. Lett.} \textbf{\bibinfo{volume}{71}}, \bibinfo{pages}{2066}
  (\bibinfo{year}{1993})\relax
\mciteBstWouldAddEndPuncttrue
\mciteSetBstMidEndSepPunct{\mcitedefaultmidpunct}
{\mcitedefaultendpunct}{\mcitedefaultseppunct}\relax
\EndOfBibitem
\bibitem[{\citenamefont{Kawasaki}(1994)}]{kawasaki1994}
\bibinfo{author}{\bibfnamefont{K.}~\bibnamefont{Kawasaki}},
  \bibinfo{journal}{Physica A} \textbf{\bibinfo{volume}{208}},
  \bibinfo{pages}{35} (\bibinfo{year}{1994})\relax
\mciteBstWouldAddEndPuncttrue
\mciteSetBstMidEndSepPunct{\mcitedefaultmidpunct}
{\mcitedefaultendpunct}{\mcitedefaultseppunct}\relax
\EndOfBibitem
\bibitem[{\citenamefont{Kawasaki and Miyazima}(1997)}]{kawasaki1997}
\bibinfo{author}{\bibfnamefont{K.}~\bibnamefont{Kawasaki}} \bibnamefont{and}
  \bibinfo{author}{\bibfnamefont{S.}~\bibnamefont{Miyazima}},
  \bibinfo{journal}{Z. Phys. B} \textbf{\bibinfo{volume}{103}},
  \bibinfo{pages}{423} (\bibinfo{year}{1997})\relax
\mciteBstWouldAddEndPuncttrue
\mciteSetBstMidEndSepPunct{\mcitedefaultmidpunct}
{\mcitedefaultendpunct}{\mcitedefaultseppunct}\relax
\EndOfBibitem
\bibitem[{\citenamefont{Yamamoto and Onuki}(1998)}]{yamamoto1998}
\bibinfo{author}{\bibfnamefont{R.}~\bibnamefont{Yamamoto}} \bibnamefont{and}
  \bibinfo{author}{\bibfnamefont{A.}~\bibnamefont{Onuki}},
  \bibinfo{journal}{Phys. Rev. Lett.} \textbf{\bibinfo{volume}{81}},
  \bibinfo{pages}{4915} (\bibinfo{year}{1998})\relax
\mciteBstWouldAddEndPuncttrue
\mciteSetBstMidEndSepPunct{\mcitedefaultmidpunct}
{\mcitedefaultendpunct}{\mcitedefaultseppunct}\relax
\EndOfBibitem
\bibitem[{\citenamefont{Fuchizaki and Kawasaki}(1998)}]{fuchizaki1998}
\bibinfo{author}{\bibfnamefont{K.}~\bibnamefont{Fuchizaki}} \bibnamefont{and}
  \bibinfo{author}{\bibfnamefont{K.}~\bibnamefont{Kawasaki}},
  \bibinfo{journal}{J. Phys. Soc. Jpn.} \textbf{\bibinfo{volume}{67}},
  \bibinfo{pages}{1505} (\bibinfo{year}{1998})\relax
\mciteBstWouldAddEndPuncttrue
\mciteSetBstMidEndSepPunct{\mcitedefaultmidpunct}
{\mcitedefaultendpunct}{\mcitedefaultseppunct}\relax
\EndOfBibitem
\bibitem[{\citenamefont{Kawasaki}(1998)}]{kawasaki1998}
\bibinfo{author}{\bibfnamefont{K.}~\bibnamefont{Kawasaki}},
  \bibinfo{journal}{J. Stat. Phys.} \textbf{\bibinfo{volume}{93}},
  \bibinfo{pages}{527} (\bibinfo{year}{1998})\relax
\mciteBstWouldAddEndPuncttrue
\mciteSetBstMidEndSepPunct{\mcitedefaultmidpunct}
{\mcitedefaultendpunct}{\mcitedefaultseppunct}\relax
\EndOfBibitem
\bibitem[{\citenamefont{Kawasaki}(2000)}]{kawasaki2000}
\bibinfo{author}{\bibfnamefont{K.}~\bibnamefont{Kawasaki}},
  \bibinfo{journal}{J. Phys.: Condens. Matter} \textbf{\bibinfo{volume}{12}},
  \bibinfo{pages}{6343} (\bibinfo{year}{2000})\relax
\mciteBstWouldAddEndPuncttrue
\mciteSetBstMidEndSepPunct{\mcitedefaultmidpunct}
{\mcitedefaultendpunct}{\mcitedefaultseppunct}\relax
\EndOfBibitem
\bibitem[{\citenamefont{Marconi and Tarazona}(1999)}]{marconi1999}
\bibinfo{author}{\bibfnamefont{U.~M.~B.} \bibnamefont{Marconi}}
  \bibnamefont{and} \bibinfo{author}{\bibfnamefont{P.}~\bibnamefont{Tarazona}},
  \bibinfo{journal}{J. Chem. Phys.} \textbf{\bibinfo{volume}{110}},
  \bibinfo{pages}{8032} (\bibinfo{year}{1999})\relax
\mciteBstWouldAddEndPuncttrue
\mciteSetBstMidEndSepPunct{\mcitedefaultmidpunct}
{\mcitedefaultendpunct}{\mcitedefaultseppunct}\relax
\EndOfBibitem
\bibitem[{\citenamefont{Marconi and Tarazona}(2000)}]{marconi2000}
  \bibinfo{journal}{J. Phys. A} \textbf{\bibinfo{volume}{12}},
  \bibinfo{pages}{413} (\bibinfo{year}{2000})\relax
\mciteBstWouldAddEndPuncttrue
\mciteSetBstMidEndSepPunct{\mcitedefaultmidpunct}
{\mcitedefaultendpunct}{\mcitedefaultseppunct}\relax
\EndOfBibitem
\bibitem[{\citenamefont{Franz and Paris}(2000)}]{franz2000}
\bibinfo{author}{\bibfnamefont{S.}~\bibnamefont{Franz}} \bibnamefont{and}
  \bibinfo{author}{\bibfnamefont{G.}~\bibnamefont{Paris}}, \bibinfo{journal}{J.
  Phys.: Condens. Matter} \textbf{\bibinfo{volume}{12}}, \bibinfo{pages}{6335}
  (\bibinfo{year}{2000})\relax
\mciteBstWouldAddEndPuncttrue
\mciteSetBstMidEndSepPunct{\mcitedefaultmidpunct}
{\mcitedefaultendpunct}{\mcitedefaultseppunct}\relax
\EndOfBibitem
\bibitem[{\citenamefont{Szamel}(2003)}]{szamel2003}
\bibinfo{author}{\bibfnamefont{G.}~\bibnamefont{Szamel}},
  \bibinfo{journal}{Phys. Rev. Lett.} \textbf{\bibinfo{volume}{90}},
  \bibinfo{pages}{228301} (\bibinfo{year}{2003})\relax
\mciteBstWouldAddEndPuncttrue
\mciteSetBstMidEndSepPunct{\mcitedefaultmidpunct}
{\mcitedefaultendpunct}{\mcitedefaultseppunct}\relax
\EndOfBibitem
\bibitem[{\citenamefont{Wu and Cao}(2005)}]{wu2005}
\bibinfo{author}{\bibfnamefont{J.}~\bibnamefont{Wu}} \bibnamefont{and}
  \bibinfo{author}{\bibfnamefont{J.}~\bibnamefont{Cao}},
  \bibinfo{journal}{Phys. Rev. Lett.} \textbf{\bibinfo{volume}{95}},
  \bibinfo{pages}{078301} (\bibinfo{year}{2005})\relax
\mciteBstWouldAddEndPuncttrue
\mciteSetBstMidEndSepPunct{\mcitedefaultmidpunct}
{\mcitedefaultendpunct}{\mcitedefaultseppunct}\relax
\EndOfBibitem
\bibitem[{\citenamefont{Miyazaki and Reichman}(2005)}]{miyazaki2005}
\bibinfo{author}{\bibfnamefont{K.}~\bibnamefont{Miyazaki}} \bibnamefont{and}
  \bibinfo{author}{\bibfnamefont{D.~R.} \bibnamefont{Reichman}},
  \bibinfo{journal}{J. Phys. A} \textbf{\bibinfo{volume}{38}},
  \bibinfo{pages}{L343} (\bibinfo{year}{2005})\relax
\mciteBstWouldAddEndPuncttrue
\mciteSetBstMidEndSepPunct{\mcitedefaultmidpunct}
{\mcitedefaultendpunct}{\mcitedefaultseppunct}\relax
\EndOfBibitem
\bibitem[{\citenamefont{Mayer et~al.}(2006)\citenamefont{Mayer, Miyazaki, and
  Reichman}}]{mayer2006}
\bibinfo{author}{\bibfnamefont{P.}~\bibnamefont{Mayer}},
  \bibinfo{author}{\bibfnamefont{K.}~\bibnamefont{Miyazaki}}, \bibnamefont{and}
  \bibinfo{author}{\bibfnamefont{D.~R.} \bibnamefont{Reichman}},
  \bibinfo{journal}{Phys. Rev. Lett.} \textbf{\bibinfo{volume}{97}},
  \bibinfo{pages}{095702} (\bibinfo{year}{2006})\relax
\mciteBstWouldAddEndPuncttrue
\mciteSetBstMidEndSepPunct{\mcitedefaultmidpunct}
{\mcitedefaultendpunct}{\mcitedefaultseppunct}\relax
\EndOfBibitem
\bibitem[{\citenamefont{Cates and Ramaswamy}(2006)}]{cates2006}
\bibinfo{author}{\bibfnamefont{M.~E.} \bibnamefont{Cates}} \bibnamefont{and}
  \bibinfo{author}{\bibfnamefont{S.}~\bibnamefont{Ramaswamy}},
  \bibinfo{journal}{Phys. Rev. Lett.} \textbf{\bibinfo{volume}{96}},
  \bibinfo{pages}{135701} (\bibinfo{year}{2006})\relax
\mciteBstWouldAddEndPuncttrue
\mciteSetBstMidEndSepPunct{\mcitedefaultmidpunct}
{\mcitedefaultendpunct}{\mcitedefaultseppunct}\relax
\EndOfBibitem
\bibitem[{\citenamefont{Andreanov et~al.}(2006)\citenamefont{Andreanov, Biroli,
  and Lef{\`{e}}vre}}]{abl}
\bibinfo{author}{\bibfnamefont{A.}~\bibnamefont{Andreanov}},
  \bibinfo{author}{\bibfnamefont{G.}~\bibnamefont{Biroli}}, \bibnamefont{and}
  \bibinfo{author}{\bibfnamefont{A.}~\bibnamefont{Lef{\`{e}}vre}},
  \bibinfo{journal}{J. Stat. Mech.} \bibinfo{pages}{P07008}
  (\bibinfo{year}{2006})\relax
\mciteBstWouldAddEndPuncttrue
\mciteSetBstMidEndSepPunct{\mcitedefaultmidpunct}
{\mcitedefaultendpunct}{\mcitedefaultseppunct}\relax
\EndOfBibitem
\bibitem[{\citenamefont{Mazenko}({\natexlab{a}})}]{mazenko2006}
\bibinfo{author}{\bibfnamefont{G.}~\bibnamefont{Mazenko}},
  \eprint{cond-mat/0609591}\relax
\mciteBstWouldAddEndPuncttrue
\mciteSetBstMidEndSepPunct{\mcitedefaultmidpunct}
{\mcitedefaultendpunct}{\mcitedefaultseppunct}\relax
\EndOfBibitem
\bibitem[{\citenamefont{Mazenko}({\natexlab{b}})}]{mazenko2007}
  \eprint{arXiv:0710.0641}\relax
\mciteBstWouldAddEndPuncttrue
\mciteSetBstMidEndSepPunct{\mcitedefaultmidpunct}
{\mcitedefaultendpunct}{\mcitedefaultseppunct}\relax
\EndOfBibitem
\bibitem[{\citenamefont{Biroli et~al.}(2006)\citenamefont{Biroli, Bouchaud,
  Miyazaki, and Reichman}}]{biroli2006}
\bibinfo{author}{\bibfnamefont{G.}~\bibnamefont{Biroli}},
  \bibinfo{author}{\bibfnamefont{J.-P.} \bibnamefont{Bouchaud}},
  \bibinfo{author}{\bibfnamefont{K.}~\bibnamefont{Miyazaki}}, \bibnamefont{and}
  \bibinfo{author}{\bibfnamefont{D.~R.} \bibnamefont{Reichman}},
  \bibinfo{journal}{Phys. Rev. Lett.} \textbf{\bibinfo{volume}{97}},
  \bibinfo{pages}{195701} (\bibinfo{year}{2006})\relax
\mciteBstWouldAddEndPuncttrue
\mciteSetBstMidEndSepPunct{\mcitedefaultmidpunct}
{\mcitedefaultendpunct}{\mcitedefaultseppunct}\relax
\EndOfBibitem
\bibitem[{\citenamefont{Biroli and Bouchaud}(2007)}]{biroli2007}
\bibinfo{author}{\bibfnamefont{G.}~\bibnamefont{Biroli}} \bibnamefont{and}
  \bibinfo{author}{\bibfnamefont{J.-P.} \bibnamefont{Bouchaud}},
  \bibinfo{journal}{J. Phys.: Condens. Matter} \textbf{\bibinfo{volume}{19}},
  \bibinfo{pages}{205101} (\bibinfo{year}{2007})\relax
\mciteBstWouldAddEndPuncttrue
\mciteSetBstMidEndSepPunct{\mcitedefaultmidpunct}
{\mcitedefaultendpunct}{\mcitedefaultseppunct}\relax
\EndOfBibitem
\bibitem[{\citenamefont{Berthier
  et~al.}(2007{\natexlab{a}})\citenamefont{Berthier, Biroli, Bouchaud, Kob,
  Miyazaki, and Reichman}}]{beither2007}
\bibinfo{author}{\bibfnamefont{L.}~\bibnamefont{Berthier}},
  \bibinfo{author}{\bibfnamefont{G.}~\bibnamefont{Biroli}},
  \bibinfo{author}{\bibfnamefont{J.-P.} \bibnamefont{Bouchaud}},
  \bibinfo{author}{\bibfnamefont{W.}~\bibnamefont{Kob}},
  \bibinfo{author}{\bibfnamefont{K.}~\bibnamefont{Miyazaki}}, \bibnamefont{and}
  \bibinfo{author}{\bibfnamefont{D.~R.} \bibnamefont{Reichman}},
  \bibinfo{journal}{J. Chem. Phys.} \textbf{\bibinfo{volume}{126}},
  \bibinfo{pages}{184503} (\bibinfo{year}{2007}{\natexlab{a}})\relax
\mciteBstWouldAddEndPuncttrue
\mciteSetBstMidEndSepPunct{\mcitedefaultmidpunct}
{\mcitedefaultendpunct}{\mcitedefaultseppunct}\relax
\EndOfBibitem
\bibitem[{\citenamefont{Berthier
  et~al.}(2007{\natexlab{b}})\citenamefont{Berthier, Biroli, Bouchaud, Kob,
  Miyazaki, and Reichman}}]{beither2007-2}
 \textbf{\bibinfo{volume}{126}},
  \bibinfo{pages}{184504} (\bibinfo{year}{2007}{\natexlab{b}})\relax
\mciteBstWouldAddEndPuncttrue
\mciteSetBstMidEndSepPunct{\mcitedefaultmidpunct}
{\mcitedefaultendpunct}{\mcitedefaultseppunct}\relax
\EndOfBibitem
\bibitem[{\citenamefont{Kim and Kawasaki}(2007)}]{kk}
\bibinfo{author}{\bibfnamefont{B.}~\bibnamefont{Kim}} \bibnamefont{and}
  \bibinfo{author}{\bibfnamefont{K.}~\bibnamefont{Kawasaki}},
  \bibinfo{journal}{J. Phys. A: Math. Theor.} \textbf{\bibinfo{volume}{40}},
  \bibinfo{pages}{F33} (\bibinfo{year}{2007})\relax
\mciteBstWouldAddEndPuncttrue
\mciteSetBstMidEndSepPunct{\mcitedefaultmidpunct}
{\mcitedefaultendpunct}{\mcitedefaultseppunct}\relax
\EndOfBibitem
\bibitem[{\citenamefont{Kim and Kawasaki}(2008)}]{kk2}
  \bibinfo{journal}{J. Stat. Mech.} \bibinfo{pages}{P02004}
  (\bibinfo{year}{2008})\relax
\mciteBstWouldAddEndPuncttrue
\mciteSetBstMidEndSepPunct{\mcitedefaultmidpunct}
{\mcitedefaultendpunct}{\mcitedefaultseppunct}\relax
\EndOfBibitem
\bibitem[{\citenamefont{Szamel}(2007)}]{szamel2007}
\bibinfo{author}{\bibfnamefont{G.}~\bibnamefont{Szamel}}, \bibinfo{journal}{J.
  Chem. Phys.} \textbf{\bibinfo{volume}{127}}, \bibinfo{pages}{084515}
  (\bibinfo{year}{2007})\relax
\mciteBstWouldAddEndPuncttrue
\mciteSetBstMidEndSepPunct{\mcitedefaultmidpunct}
{\mcitedefaultendpunct}{\mcitedefaultseppunct}\relax
\EndOfBibitem
\bibitem[{\citenamefont{Martin et~al.}(1973)\citenamefont{Martin, Siggia, and
  Rose}}]{msr}
\bibinfo{author}{\bibfnamefont{C.~P.} \bibnamefont{Martin}},
  \bibinfo{author}{\bibfnamefont{E.~D.} \bibnamefont{Siggia}},
  \bibnamefont{and} \bibinfo{author}{\bibfnamefont{H.~A.} \bibnamefont{Rose}},
  \bibinfo{journal}{Phys. Rev. A} \textbf{\bibinfo{volume}{8}},
  \bibinfo{pages}{423} (\bibinfo{year}{1973})\relax
\mciteBstWouldAddEndPuncttrue
\mciteSetBstMidEndSepPunct{\mcitedefaultmidpunct}
{\mcitedefaultendpunct}{\mcitedefaultseppunct}\relax
\EndOfBibitem
\bibitem[{\citenamefont{Dean}(1996)}]{dean1996}
\bibinfo{author}{\bibfnamefont{D.}~\bibnamefont{Dean}}, \bibinfo{journal}{J.
  Phys. A} \textbf{\bibinfo{volume}{29}}, \bibinfo{pages}{L613}
  (\bibinfo{year}{1996})\relax
\mciteBstWouldAddEndPuncttrue
\mciteSetBstMidEndSepPunct{\mcitedefaultmidpunct}
{\mcitedefaultendpunct}{\mcitedefaultseppunct}\relax
\EndOfBibitem
\bibitem[{\citenamefont{Das and Mazenko}()}]{dm2}
\bibinfo{author}{\bibfnamefont{S.~P.} \bibnamefont{Das}} \bibnamefont{and}
  \bibinfo{author}{\bibfnamefont{G.}~\bibnamefont{Mazenko}},
  \eprint{arXiv:0801.1727}\relax
\mciteBstWouldAddEndPuncttrue
\mciteSetBstMidEndSepPunct{\mcitedefaultmidpunct}
{\mcitedefaultendpunct}{\mcitedefaultseppunct}\relax
\EndOfBibitem
\bibitem[{\citenamefont{Kim and Mazenko}(1991)}]{kim1991}
\bibinfo{author}{\bibfnamefont{B.}~\bibnamefont{Kim}} \bibnamefont{and}
  \bibinfo{author}{\bibfnamefont{G.~F.} \bibnamefont{Mazenko}},
  \bibinfo{journal}{J. Stat. Phys.} \textbf{\bibinfo{volume}{64}},
  \bibinfo{pages}{631} (\bibinfo{year}{1991})\relax
\mciteBstWouldAddEndPuncttrue
\mciteSetBstMidEndSepPunct{\mcitedefaultmidpunct}
{\mcitedefaultendpunct}{\mcitedefaultseppunct}\relax
\EndOfBibitem
\bibitem[{\citenamefont{Hansen and McDonald}(1986)}]{hansen}
\bibinfo{author}{\bibfnamefont{J.~P.} \bibnamefont{Hansen}} \bibnamefont{and}
  \bibinfo{author}{\bibfnamefont{I.~R.} \bibnamefont{McDonald}}, in
  \emph{\bibinfo{booktitle}{Theory of simple liquids}}
  (\bibinfo{publisher}{Academic Press}, \bibinfo{address}{London},
  \bibinfo{year}{1986})\relax
\mciteBstWouldAddEndPuncttrue
\mciteSetBstMidEndSepPunct{\mcitedefaultmidpunct}
{\mcitedefaultendpunct}{\mcitedefaultseppunct}\relax
\EndOfBibitem
\bibitem[{\citenamefont{Ramakrishnan and
  Yussouff}(1979)}]{ramakrishnan_yussouff}
\bibinfo{author}{\bibfnamefont{T.~V.} \bibnamefont{Ramakrishnan}}
  \bibnamefont{and} \bibinfo{author}{\bibfnamefont{M.}~\bibnamefont{Yussouff}},
  \bibinfo{journal}{Phys. Rev. B} \textbf{\bibinfo{volume}{19}},
  \bibinfo{pages}{2775} (\bibinfo{year}{1979})\relax
\mciteBstWouldAddEndPuncttrue
\mciteSetBstMidEndSepPunct{\mcitedefaultmidpunct}
{\mcitedefaultendpunct}{\mcitedefaultseppunct}\relax
\EndOfBibitem
\bibitem[{\citenamefont{Jensen}(1981)}]{jensen1981}
\bibinfo{author}{\bibfnamefont{R.~V.} \bibnamefont{Jensen}},
  \bibinfo{journal}{J. Stat. Phys.} \textbf{\bibinfo{volume}{25}},
  \bibinfo{pages}{183} (\bibinfo{year}{1981})\relax
\mciteBstWouldAddEndPuncttrue
\mciteSetBstMidEndSepPunct{\mcitedefaultmidpunct}
{\mcitedefaultendpunct}{\mcitedefaultseppunct}\relax
\EndOfBibitem
\bibitem[{\citenamefont{Lust et~al.}(1993)\citenamefont{Lust, Valls, and
  Dasgupta}}]{lust1993}
\bibinfo{author}{\bibfnamefont{L.~M.} \bibnamefont{Lust}},
  \bibinfo{author}{\bibfnamefont{O.~T.} \bibnamefont{Valls}}, \bibnamefont{and}
  \bibinfo{author}{\bibfnamefont{C.}~\bibnamefont{Dasgupta}},
  \bibinfo{journal}{Phys. Rev. E} \textbf{\bibinfo{volume}{48}},
  \bibinfo{pages}{1787} (\bibinfo{year}{1993})\relax
\mciteBstWouldAddEndPuncttrue
\mciteSetBstMidEndSepPunct{\mcitedefaultmidpunct}
{\mcitedefaultendpunct}{\mcitedefaultseppunct}\relax
\EndOfBibitem
\end{mcitethebibliography}

\end{document}